\documentclass[reprint,prd,nofootinbib,superscriptaddress,showpacs]{revtex4-1}
\usepackage[mathcal]{euscript}
\usepackage[latin1]{inputenc}
\usepackage{latexsym}
\usepackage{amsmath}
\usepackage{hyperref}
\usepackage{amsmath}    
\usepackage{graphicx}   
\usepackage{verbatim}   
\usepackage{color}      
\usepackage{subfigure}  
\usepackage{hyperref}   
\usepackage{bm}
\usepackage{graphicx}
\usepackage{amssymb}
\usepackage{bbm}
\usepackage{float}
\usepackage{slashed}
\usepackage{amsfonts}
\usepackage{euscript}
\usepackage{amsmath}    
\usepackage{mathrsfs} 
\usepackage{bbding}
\usepackage{pifont}
\newcommand{\cm}[1]{}
\usepackage{cancel}
\makeindex

\begin{document}

\title{Unimodular gravity and general relativity from graviton self-interactions}

\author{Carlos Barcel\'o}
\email{carlos@iaa.es}
\affiliation{Instituto de Astrof\'{\i}sica de Andaluc\'{\i}a (IAA-CSIC), Glorieta de la Astronom\'{\i}a, 18008 Granada, Spain}
\author{Ra\'ul Carballo-Rubio}
\email{raulc@iaa.es}
\affiliation{Instituto de Astrof\'{\i}sica de Andaluc\'{\i}a (IAA-CSIC), Glorieta de la Astronom\'{\i}a, 18008 Granada, Spain}
\author{Luis J. Garay}
\email{luisj.garay@fis.ucm.es}
\affiliation{Departamento de F\'{\i}sica Te\'orica II, Universidad Complutense de Madrid, 28040 Madrid, Spain}
\affiliation{Instituto de Estructura de la Materia (IEM-CSIC), Serrano 121, 28006 Madrid, Spain}

\begin{abstract}{

It is commonly accepted that general relativity is the only solution to the consistency problem that appears when trying to build a theory of interacting gravitons (massless spin-2 particles). Padmanabhan's 2008 thought-provoking analysis raised some concerns that are having resonance in the community. In this work we present the self-coupling problem in detail and explicitly solve the infinite-iterations scheme associated with it for the simplest theory of a graviton field, which corresponds to an irreducible spin-2 representation of the Poincar\'e group. We make explicit the non-uniqueness problem by finding an entire family of solutions to the self-coupling problem. Then we show that the only resulting theory which implements a deformation of the original gauge symmetry happens to have essentially the structure of unimodular gravity. This makes plausible the possibility of a natural solution to the first cosmological constant problem in theories of emergent gravity. Later on we change for the sake of completeness the starting free-field theory to Fierz-Pauli theory, an equivalent theory but with a larger gauge symmetry. We indicate how to carry out the infinite summation procedure in a similar way. Overall, we conclude that as long as one requires the (deformed) preservation of internal gauge invariance, one naturally recovers the structure of unimodular gravity or general relativity but in a version that explicitly shows the underlying Minkowski spacetime, in the spirit of Rosen's flat-background bimetric theory.

}
\end{abstract}
\keywords{graviton, self-coupling, cosmological constant, vacuum energy, unimodular gravity, emergent gravity}
\pacs{04.20.-q, 04.60.-m, 04.62.+v, 11.30.Cp}
\maketitle
\flushbottom
%

\section{Introduction}

The tight connection between general relativity and geometry is what makes this theory conceptually beautiful, but also very different from the formalism developed to describe the other fundamental forces we know about: the standard model of particle physics. While the latter is formulated as a quantum field theory in Minkowski spacetime, in general relativity there is no such a notion of preferred, immutable arena in which physics takes place. Instead this environment (spacetime) is also a dynamical object in its own right. This is arguably the root of the conceptual problems concerning the reconciliation between general relativity and quantum mechanics.

Trying to bridge this gap Rosen \cite{Rosen1940,Rosen1940b} showed that general relativity can be reinterpreted as a nonlinear field theory over Minkowski spacetime. Later Gupta \cite{Gupta1954} proposed that a consistent theory of self-interacting gravitons should have precisely the structure of general relativity. In brief, Gupta's idea is to start with a free massless spin-2 field in Minkowski spacetime, and then make it interact with the rest of fields. General considerations show that the fact that this field has spin 2 implies that this can be done only if the graviton field interacts with itself, making the overall theory nonlinear. Since one can always express general relativity plus matter as a nonlinear theory for the deviations of the metric with respect to some flat reference metric, there is wiggle room to reconcile both visions.

However, to make this programme come to fruition, one should be able to determine the nature of the resulting nonlinear theory which arises from the self-coupling of the graviton field. The first subtle point is that the Lagrangian density of such a theory contains, in principle, infinite interaction terms which are obtained consecutively by an iterative process, so one should devise a way to manage them and show that this infinite series converges to the Lagrangian density of general relativity. This question was indirectly addressed in the work of Kraichnan \cite{Kraichnan1955} and Feynman \cite{Feynman2002}, but was finally settled by Deser \cite{Deser1970}. To do that, he used specific variables which make the series finite, thus avoiding to perform the sum of an infinite series.

The second source of concern is the non-uniqueness of the construction as there are many and, in principle, inequivalent ways to make the graviton field self-interact. This was first raised by Huggins in his 1962 thesis \cite{Huggins1962}. The central point of his argument is that one needs more information to uniquely fix the stress-energy tensor of the graviton field to which it couples itself. Thus there are potentially many self-interacting theories and, as there is no control of those theories, it is not easy to conclude whether or not they are equivalent to general relativity. Recently, Padmanabhan has raised equivalent arguments \cite{Padmanabhan2008}. In fact, this work has been motivated by Padmanabhan's paper, a subsequent follow up by Butcher et al.~\cite{Butcher2009}, and the reply by Deser~\cite{Deser2010}. 
 
In this work we start by considering the simplest theory of a graviton field, corresponding to an irreducible spin-2 representation of the Poincar\'e group. After describing the linear spin-2 theory, we develop in detail the self-interacting scheme and find the (formal) sum of the series for the action. To the best of our knowledge, it is the first time that an infinite series arising in the graviton self-coupling problem is summed. This construction is not unique. We present the non-uniqueness problem and show that, indeed, there is a one-parameter family of solutions to the self-coupling problem which were not found in previous approaches. We analyze these resulting theories and show that requiring the preservation of gauge symmetry can be used to single out one of them, which turns out to be equivalent to unimodular gravity. Ater the detailed analysis of the spin-2 theory, we move to compare this approach with the more standard that starts with Fierz-Pauli theory. We sketch how to solve in an equivalent way the self-coupling problem in this case. We study in a similar way the issue of uniqueness and then, at the light of our results, we present a discussion aimed at reconciling what seem disparate results in the recent literature \cite{Padmanabhan2008,Butcher2009,Deser2010}.

On the one hand our results confirm the concerns of \cite{Padmanabhan2008} in that the self-coupling problem by itself does not uniquely lead to unimodular gravity (or general relativity, depending on the starting linear theory) unless further conditions are imposed along the process. Specifically, one needs to require that the gauge structure of the initial linear theory is preserved, although deformed, in the final outcome. This condition singles out unimodular gravity (or general relativity) but in a version that explicitly shows the underlying Minkowski spacetime, in the spirit of Rosen's flat-background bimetric theory. On the other hand, once the gauge preservation condition is applied, the entire construction can be taken to completion in a natural way using only flat spacetime notions, position which is defended in \cite{Deser2010}. In fact, the presence of the Minkowski background structure permits to clearly separate the internal gauge transformations from invariance under changes of coordinates. Within this bimetric construction one obtains a quadratic Lagrangian density, invariant under changes of coordinates. However, this quadratic Lagrangian density is not invariant under internal gauge transformations, but the action is. One could take a further leap and adopt a geometrical (single-metric) interpretation \cite{Deser1970,Deser2010}. From this perspective the quadratic Lagrangian density is not diffeomorphism invariant: the surface term of the Einstein-Hilbert action cannot be recovered and would have to be added by hand, as pointed out in \cite{Padmanabhan2008}. Although certainly appealing, within the self-coupling problem we do not find a compelling reason to take this geometrization leap.

The structure of the paper is the following. Sections \ref{sec:free} and \ref{sec:matter} are intended to be a recapitulation of the knowledge which can be found in different sources, concerning the field-theoretical description of gravitons and the consistency problem which appears when one tries to make them interact with matter. This problem leads to the consideration of a self-interacting scheme for the gravitons, which is developed in the next two sections, the core of the paper. In section \ref{sec:selfc} we consider the simplest theory of a graviton field, corresponding to an irreducible spin-2 representation of the Poincar\'e group. We develop the self-interacting scheme and find the (formal) sum of the series for the action. In section \ref{sec:full} we discuss how the condition of preservation of the maximal amount of internal gauge invariance can eliminate the ambiguities inherent to the self-coupling procedure. We also show that this outcome of the self-coupling problem is equivalent to unimodular gravity. Section \ref{sec:prev} is devoted to show how to apply the same programme to Fierz-Pauli theory and how general relativity comes into play. We also include a discussion on gravitational energy which is particularly interesting in the case of unimodular gravity. We end with a brief summary and some conclusions.

Notation and writing style: we use the metric convention $(-,+,+,+)$ and we will always avoid making explicit the spatiotemporal dependence of the different fields considered in the text. No distinction is made between Greek and Latin indices. By gravitons we mean massless spin-2 particles, though this notion still has wiggle room to permit different implementations. Here we are considering two types of graviton fields: the spin-2 and Fierz-Pauli fields. When the Minkowski metric $\eta_{ab}$ is used, it is understood as  written in a generic coordinate system. The d'Alembert operator in the flat metric is $\square$. Similarly, the covariant derivative $\nabla$ is always related to the flat metric, while $\nabla'$ corresponds to a curved metric. Curvature-related quantities will be defined by following Misner-Thorne-Wheeler's convention \cite{Wheeler1973}. This work was initially motivated by the paper of Padmanabhan \cite{Padmanabhan2008} and, thus, we have decided to partially maintain his notation to facilitate the translation of the results. Our intention throughout the paper has been to use an ``aseptic'' writing style that avoids contamination from geometric notions motivated by previous knowledge of general relativity. In this way, one can see more clearly the different steps which are necessary to obtain general relativity (or unimodular gravity) as solutions to the self-coupling problem.

\section{Free graviton fields \label{sec:free}}

The unitary representations of the Poincar\'e group as first classified by Wigner are determined by the value of the mass $m$ and the eigenvalues of the so-called little group \cite{Wigner1939,Loebbert2008}. For a particle with mass $m\neq0$ the little group is $\text{SO}(3)$, so the corresponding label is the angular momentum $j$ and one has $2j+1$ states in each representation, corresponding to polarizations which range from $\sigma=-j$ to $\sigma=+j$ jumping in units. However, for a massless particle the little group is $\text{ISO}(2)$ (the 2-dimensional Euclidean group) and only the states with polarizations $\sigma=\pm j$ are left. This means that massless particles with integer spin carry only two independent degrees of freedom. As linear representation space one would like to construct a tensor-field space using exclusively these degrees of freedom, but it is in this construction where gauge invariance appears inevitably intertwined with Lorentz invariance. In the following, we are going to work in a covariant fashion with respect to changes of coordinates in Minkowski spacetime. This will be useful to distinguish between covariance and internal gauge invariance in the resulting nonlinear theory.

The natural way to define such a spacetime tensor field is to construct an object $\frak{h}^{ab}$ exclusively made of the two physical polarizations, which can be done for example in momentum space \cite{Weinberg2000}. The problem is that Lorentz transformations do not leave this space invariant, not even the transformations that belong to the little group. Assuming a tensorial character for $\frak{h}^{ab}$ we have that, under an infinitesimal Lorentz transformation whose generators are $\xi^a_\omega$, it transforms as 
\begin{equation}
h_\omega^{ab} = \frak{h}^{ab}+ \xi_\omega^c \nabla_c \frak{h}^{ab}+\frak{h}^{ac}\nabla_c\xi_\omega^b+\frak{h}^{bc}\nabla_c\xi_\omega^a.\label{eq:lortrans}
\end{equation}
The Lorentz generators $\xi_\omega$ verify the conditions
\begin{equation}
\nabla_a\xi_\omega^a=0,\qquad \square \xi_\omega^a=0.\label{eq:lgen}
\end{equation}
The problem is that the last terms in this transformation law do not belong to the linear space of objects of $\frak{h}^{ab}$-type~\cite{Jenkins2006}, i.e.
\begin{equation}
\xi_\omega^c \nabla_c \frak{h}^{ab}+\frak{h}^{ac}\nabla_c\xi_\omega^b+\frak{h}^{bc}\nabla_c\xi_\omega^a
\neq {\frak{h}'}^{ab}
.\label{eq:noninv}
\end{equation}
Thus, the transformed object $h_\omega^{ab}$ does not belong to the linear representation space we started with (the same problem appears in electrodynamics, where the gauge fixing condition $\epsilon^0=0$ on the polarization vector $\epsilon^\mu$ is not Lorentz-invariant). One can prove instead that the terms driving one outside the representation space are of the form
\begin{equation}
h_\omega^{ab} -{\frak{h}'}_\omega^{ab}= \eta^{ac}\nabla_c\xi_\omega^b+\eta^{bc}\nabla_c\xi_\omega^a.\label{eq:lortrans}
\end{equation}

To circumvent the representation problem we have two options: instead of using a tensor space as representation space one could define $\frak{h}^{ab}$ as a non-tensorial object; the other possibility is to maintain a subsidiary tensor-field space but consider as representation space not its individual elements but equivalence classes of them related by gauge transformations. We shall proceed using this second approach.

The minimal realization of the gauge approach is to take as representation space tensorial objects $h^{ab}$ such that they are traceless and transverse,
\begin{equation}
\eta_{ab}h^{ab}=0,\qquad \nabla_bh^{ab}=0,\label{eq:spin2cons}
\end{equation}
and with equation of motion 
\begin{equation}
\square h^{ab}=0.\label{eq:freeeqmot}
\end{equation}
Moreover, any $h^{ab}$ and ${h'}^{ab}$ related by an internal gauge transformation,
\begin{equation}
h'^{ab} = h^{ab}+\eta^{ac}\nabla_c\xi^b+\eta^{bc}\nabla_c\xi^a,\label{eq:freegauge}
\end{equation}
will represent the same physical configuration, with generators verifying the conditions
\begin{equation}
\nabla_a\xi^a=0,\qquad \square \xi^a=0.\label{eq:tdiff}
\end{equation}
Notice that the internal gauge transformation (\ref{eq:freegauge}) and the external transformation associated with a general change of coordinates with generators $\tilde{\xi}^a$,
\begin{equation}
{h'}^{ab} = h^{ab}+ \tilde{\xi}^c \nabla_c h^{ab}+
h^{ac}\nabla_c\tilde{\xi}^b+h^{bc}\nabla_c\tilde{\xi}^a,\label{eq:gentrans}
\end{equation}
are completely different from each other: the space of generators is different and so is their implementation in the symmetry transformation. Moreover, the last transformation affects the coordinates and the rest of fields.

It is easy to check that the traceless and transverse conditions are preserved by these internal gauge transformations. These constraints in the definition of the field $h^{ab}$ can be thought as the elimination of the scalar and vector representations of the Poincar\'e group (see appendix I in \cite{Ogievetsky1965}). A detailed analysis shows that one can always find a vector $\xi^a$ such that the states with helicites $\sigma=0,\pm1$ are gauged away or, in other words, the corresponding components $h^{00}$ and $h^{0i}$, $i=1,2,3$ are set to zero while the remaining components are constrained so that there are two independent degrees of freedom. Another common choice to show this is the light-cone gauge \cite{Zwiebach2009}.

Notice that there are only two contractions of $(\nabla h)^2$ with metric objects ($\eta_{ab}$, $\eta^{ab}$ and $\delta^a_b$) which are not zero by the traceless and transverse conditions \eqref{eq:spin2cons}. By virtue of this, the Lagrangian density should have the form
\begin{equation}
\mathscr{L}_{\text{G},0}:=c_1\eta^{ai}\eta_{bj}\eta_{ck}\nabla_ah^{bc}\nabla_ih^{jk}+c_2\delta^a_k\delta^i_c\eta_{bj}\nabla_ah^{bc}\nabla_ih^{jk}.\label{eq:free5}
\end{equation}
Here $c_1$ and $c_2$ are real constants. The second term is equivalent to a total divergence because of the transverse condition in \eqref{eq:spin2cons}, so it does not affect the form of the equations of motion. However, its presence can affect the definition of the source of the self-interacting equations as we will see. We can set the normalization to $c_1=-1/4$ and introduce a bit of notation to conveniently write the free action as:
\begin{equation}
\mathscr{A}_{\text{G},0}:=\frac{1}{4}\int\text{d}\mathscr{V}_\eta\,M^{ai}_{\mathfrak{s} \ \ bcjk}(\eta)\nabla_a h^{bc}\nabla_i h^{jk},\label{eq:free9}
\end{equation}
where $\text{d}\mathscr{V}_\eta:=\text{d}^4x\sqrt{-\eta}$ is the Minkowski volume element and the Lorentz tensor $M^{ai}_{\mathfrak{s} \ \ bcjk}(\eta)$ is given by:
\begin{align}
M^{ai}_{\mathfrak{s} \ \ bcjk}(\eta):=\mathfrak{s}\left[\eta_{b(j}\delta^a_{k)}\delta^i_{c}+\eta_{c(j}\delta^a_{k)}\delta^i_{b}\right]-\nonumber\\
-\eta^{ai}\eta_{b(j}\eta_{k)c}.\label{eq:free10}
\end{align}
The tensorial quantity $M^{ai}_{\mathfrak{s} \ \ bcjk}(\eta)$ is symmetric under $b\leftrightarrow c$, $j\leftrightarrow k$ and $(a,b,c)\leftrightarrow(i,j,k)$, as one can check from its definition. When used in the action we do not need to worry about these symmetries because it is contracted with an object, $(\nabla h)^2$, which already has these symmetries. However, to solve the iterative equations of the self-coupling problem it will be necessary to use its symmetric form as it appears in \eqref{eq:free10}. The parameter $\mathfrak{s}$, directly proportional to $c_2$, controls the surface terms we are considering in the free action. One can easily check that this action is invariant up to a surface term under the gauge transformations \eqref{eq:freegauge} with generators satisfying \eqref{eq:tdiff}. In fact, the case $\mathfrak{s}=1$ is special in the sense that one could drop the second condition in \eqref{eq:tdiff} and these transformations are still a symmetry. For this reason we will always assume that this is the case i.e. only the first condition in \eqref{eq:tdiff} applies when considering $\mathfrak{s}=1$, thus recovering the minimal theory of gravitons which was considered in \cite{vanderBij1981}. The reader will notice that, through the calculations in section \ref{sec:selfc}, we always keep using the object $h^{ab}$ and never its covariant counterpart $h_{ab}:=\eta_{ac}\eta_{bd}h^{cd}$. This simplifies some steps which involve taking variational derivatives with respect to an auxiliary metric $\gamma_{ab}$ after the replacement $\eta_{ab}\rightarrow\gamma_{ab}$ in the action. In the following, this traceless field will be called spin-2 field.

An alternative way of constructing a spacetime field $h^{ab}$ is to drop the traceless and transverse conditions in~\eqref{eq:spin2cons} while enlarging the gauge symmetry. This is the well-known Fierz-Pauli theory \cite{FierzPauli1939}, in which the fundamental field is just a symmetric Lorentz tensor. We will call it Fierz-Pauli field. We are not going to enter into details here about the construction of the Fierz-Pauli action as the procedure is equivalent (but a little more involved) to the one we have followed for the spin-2 field. The details can be found e.g. in \cite{Padmanabhan2008}, and we will mention part of them in section \ref{sec:prev}. Notice that the transverse and traceless conditions can be imposed on the Fierz-Pauli field only within the space of solutions of the free theory. That is, the so-called transverse-traceless gauge can be applied only for fields $h^{ab}$ verifying the condition
\begin{equation}
\nabla_a\nabla_bh^{ab}=\eta_{ab}\square h^{ab},
\end{equation}
which is precisely the trace of the Fierz-Pauli dynamical equations \cite{Ortin2004}.

To develop the self-coupling scheme, we are going to treat these theories as classical field theories. Our conclusions are applicable then to the long wavelength limit of theories in which a graviton propagates over Minkowski space in interaction with matter, independently of the ultraviolet completion of the theory.\footnote{In particular, the notion of a Minkowski preferred background could be emergent in the sense of being applicable only below some characteristic energy scale, instead of a fundamental structure present in all regimes \cite{Barcelo2011}.} From the point of view of the classical equations of motion, which are what we are interested in, surface terms in the resulting action of the self-interacting theory are irrelevant. Notice that, if there is a regime in the theory in which gravity functions classically but the matter fields behave quantum-mechanically (semiclassical gravity), it is reasonable to expect that our conclusions would also apply to it as the self-coupling only occurs in the gravitational sector, still described by $c$-numbers.

\section{Coupling to matter \label{sec:matter}}

In this section we address the question: could we make an interacting theory of the spin-2 and matter fields? We have included this review section to facilitate the reading of the paper, but its contents are well known in the literature (see \cite{Ortin2004} and references therein).

To couple a matter field to the spin-2 field we need to define a quantity $\widetilde{T}_{ab}$ that is symmetric, traceless and transverse on solutions. Then, we could write:
\begin{equation}
\square h_{ab}=\lambda \widetilde{T}_{ab}.\label{eq:fans}
\end{equation}
A natural consistency condition is to impose that this equation can be obtained from an action. In fact, it can be obtained as the corresponding Euler-Lagrange equation with respect to restricted variations of $h^{ab}$ (such that $\eta_{ab}\delta h^{ab}=0$) as long as we add to the Lagrangian density a term
\begin{equation}
\Delta\mathscr{L}:=\lambda h^{ab}T_{ab},
\end{equation}
with $T_{ab}$ symmetric, transverse on solutions, and with constant trace. An object with these characteristics is precisely the stress-energy momentum tensor, where the transverse condition amounts to its conservation. We comment on the condition of constancy of the trace at the end of the section; for now let us just require conservation of $T_{ab}$.

One inmediately realizes that one cannot use the stress-energy tensor of the free matter theory: for consistency one must use the total stress-energy tensor of the interacting theory. Had we started tentatively by adding a term $\lambda h^{ab}T_{ab}^{\text{M}}$ to the Lagrangian density of the matter sector, with $T_{ab}^{\text{M}}$ the free matter stress-energy tensor, this very term would have changed the matter stress-energy tensor making necessary to add new energetic contributions to the Lagrangian density. This iterative process is the one we are going to follow in the next section.  

This is a general property of the coupling with matter: as the coupling is done through the stress-energy tensor, the transverse condition for the spin-2 field implies that the total matter traceless stress-energy tensor should be divergenceless.\footnote{It can be seen that when the transverse condition is relaxed, it is the resulting gauge invariance of the theory the responsible for this feature \cite{Ortin2004}.} However, this would not be the case when interaction is switched on, as the matter fields no longer behave as an isolated system, being the energy transferred between them and the spin-2 field. The natural way to remedy this is to realize that the spin-2 field must also act as a source of itself (the charge/source of the graviton field is the energy and gravitons should possess energy), which leads us to the issue of the spin-2 field self-coupling. Therefore, the iterative procedure has to act also in the spin-2 sector. 

An important problem shows up when thinking about the stress-energy tensor of the spin-2 sector: there is no way of constructing a non-trivial conserved stress-energy tensor for the spin-2 field that is invariant under the gauge transformations \eqref{eq:freegauge} \cite{Aragone1971,Magnano2002}. By non-trivial we mean that it is not exactly zero for any solution. An indirect way of realizing it could be the Weinberg-Witten theorem \cite{Weinberg1980,Loebbert2008}, which explicitly forbids this possibility. Thus, one cannot associate a local notion of energy with the physical configurations in the free theory.\footnote{Something equivalent happens in non-abelian Yang-Mills theory: one cannot find a Lorentz covariant conserved current which is at the same time gauge invariant.} One can live with this fact if the theory is non-interacting, so that there is no operational way to define what it is energy-momentum. Within an interaction scheme this is untenable.

As mentioned before, strictly speaking the consistency of a complete nonlinear spin-2 theory will need that the transverse condition applies to the traceless object $\widetilde{T}_{ab}$, independently of that of $T_{ab}$. When the trace of the stress-energy tensor is a constant the conservation of $T_{ab}$ implies the conservation of $\widetilde{T}_{ab}$. On this respect, notice that the trace of the stress-energy tensor of a single linear matter sector (e.g. a single scalar field) is always a constant as there are no sources that can cause an inhomogeneity in the system, and hence in the trace. When putting together the spin-2 and matter sectors with all its nonlinear interactions, we do not know a priori what could happen. All in all one is left with the expectation that when applying a consistent self-interacting scheme some meaningful result would show up for all the standing problems. We will come back to this issue after finding the solutions to the self-coupling problem. 

\section{Self-coupling \label{sec:selfc}}

The first question to answer when considering a self-interacting scheme is: what is the object to which we are going to couple the spin-2 field in a first stage? A natural candidate to consider is the canonical stress-energy tensor, that is, a conserved quantity of any field theory which is Poincar\'e invariant, associated with invariance under translations:
\begin{equation}
\Theta^a_{\ b}:=\mathscr{L}_0\delta^a_b-\frac{\delta\mathscr{L}_0}{\delta(\nabla_a \psi^\mu)}\nabla_b\psi^\mu,\qquad \nabla_a\Theta^{a}_{\ b}=0.\label{eq:sint1}
\end{equation}
Here $\mathscr{L}_0$ is the free Lagrangian density of both spin-2 and matter fields, collectively denoted by $\psi^\mu$. If we manage to use this quantity as the right-hand side of our equations of motion for $h^{ab}$, the theory will naturally verify the condition obtained by Weinberg \cite{Weinberg1964,Weinberg1964b} as a necessary one if we want to have a Lorentz-invariant theory: the coupling between the spin-2 field to matter and to itself must be governed by the same coupling constant. 

However, direct use of this quantity is not possible: in general, the fully covariant or contravariant counterparts of \eqref{eq:sint1} are not symmetric. But we can exploit the ambiguity in the addition of identically conserved tensors, the so-called Belinfante-Rosenfeld terms, to obtain a symmetric stress-energy tensor which leads to the same conserved quantities. This symmetric stress-energy tensor is not unique: one can still add identically conserved tensors keeping the symmetric character. All the manipulations that follow can be performed by directly using the symmetrized versions of the canonical stress-energy tensor. Therefore, these manipulations in no way involve any curved spacetime notion. However, as shown by Belinfante and Rosenfeld \cite{Belinfante1940,Rosenfeld1940}, these symmetric stress-energy tensors can be equivalently obtained by the simple Hilbert prescription,
\begin{equation}
T_{ab}:=-\lim_{\gamma\rightarrow\eta}\frac{1}{\sqrt{-\gamma}}\frac{\delta \mathscr{A}_0}{\delta \gamma^{ab}},\label{eq:tabdef}
\end{equation}
where the flat metric $\eta_{ab}$ in $\mathscr{A}_0$ has been replaced by an auxiliary (generally curved) metric $\gamma_{ab}$, being $\gamma^{ab}$ its inverse. Recall that the two steps one needs to follow are: write the action in curvilinear coordinates in flat space, and then generalize it to curved space. It is in this second step where the ambiguities show up in Hilbert's prescription. In practice, the ambiguities in the stress-energy tensor appear now as the addition of non-minimal couplings of the physical fields to the auxiliary metric $\gamma_{ab}$, and they are added to the one-parameter family of surface terms we are considering in the free action \eqref{eq:free9}. In fact, as we will see later these non-minimal couplings can be understood as surface terms in the original free action, though different from the one-parameter family we have been considering up to now. We will show that these different choices of stress-energy tensor as the source lead to different solutions to the self-coupling problem. Let us stress again that here we use Hilbert's prescription as a mere calculational device and insist in that no curved spacetime notion is used throughout the calculations.

Now that we have discussed the relevant properties of the stress-energy tensor, we would like to derive the self-interacting equations of motion from an action. The coupling constant is denoted by $\lambda$. This can be done if we add a term $\lambda\mathscr{A}_1$ of order $\mathscr{O}(\lambda)$ in the action, such that:
\begin{equation}
\frac{\delta\mathscr{A}_1}{\delta h^{ab}}=\lim_{\gamma\rightarrow\eta}\frac{\delta\mathscr{A}_{0}}{\delta\gamma^{ab}}.\label{eq:ford}
\end{equation}
Here $\mathscr{A}_0$ has two terms: the free action for the spin-2 field given by \eqref{eq:free9} and the matter content one wants to consider, $\mathscr{A}_0=\mathscr{A}_{\text{G},0}+\mathscr{A}_{\text{M},0}$.

As noticed by Gupta \cite{Gupta1954}, this additional term of order $\mathscr{O}(\lambda)$ in the action would modify the definition of the source by a term of order $\mathscr{O}(\lambda^2)$, which implies that we need to contemplate a term $\lambda^2\mathscr{A}_2$ in the action. This is the iterative procedure we would want to solve for. It will generate an action of the form\footnote{Notice that there is no reason to expect, in principle, this series to be infinite. There are two examples in the literature of this kind of series: the first one is the trivial one, in the sense that one does not consider self-interactions of the graviton field (see paragraph below). This series is infinite. The only example of a self-interacting series is the one constructed by Deser \cite{Deser1970} which is finite, with only $\mathscr{A}_1\neq0$. In this work we are going to consider always infinite series.}
\begin{equation}
\mathscr{A}=\sum_{n=0}^\infty \lambda^n\mathscr{A}_n,
\end{equation}
where the set of partial actions $\{\mathscr{A}_n\}_{n=1}^\infty$ must verify the iterative equations
\begin{equation}
\frac{\delta\mathscr{A}_n}{\delta h^{ab}}=\lim_{\gamma\rightarrow\eta}\frac{\delta\mathscr{A}_{n-1}}{\delta\gamma^{ab}},\qquad n\geq1.\label{eq:cons3}
\end{equation}
In more detail, let us write the resulting action of the self-coupling procedure as
\begin{equation}
\mathscr{A}:=\mathscr{A}_0+\mathscr{A}_{\text{I}},
\end{equation}
where the free part $\mathscr{A}_0$ is already defined and $\mathscr{A}_{\text{I}}$ is the self-interacting part we are going to solve for. Given this action, one would be able to obtain its stress-energy tensor. $\mathscr{A}_{\text{I}}$ is then fixed by the requeriment of leading to this stress-energy tensor as the source of the equations of motion:
\begin{equation}
\frac{\delta\mathscr{A}_{\text{I}}}{\delta h^{ab}}=\lambda\lim_{\gamma\rightarrow\eta}\frac{\delta(\mathscr{A}_0+\mathscr{A}_{\text{I}})}{\delta\gamma^{ab}}.\label{eq:app1eq1}
\end{equation}
One just needs to expand $\mathscr{A}_{\text{I}}=\sum_{n=1}^\infty\lambda^n\mathscr{A}_n$ and compare different orders in the coupling constant $\lambda$ to obtain the set of iterative equations \eqref{eq:cons3}.

Recall that we started with traceless equations of motion. This means that, to keep the same number of equations of motion in the iterative procedure, the field $h^{ab}$ must be constrained in some way. One option is to maintain the traceless condition with respect to the original Minkowski metric. In any case, the set of equations \eqref{eq:cons3} is general enough to permit the imposition of this condition, as well as other possible scalar constraints over the field $h^{ab}$, after finding its solution. So we postpone this discussion to section \ref{sec:full}, although we will keep in mind the existence of this constraint which, at least at the lowest order, must be equivalent to the traceless condition.

The integration of the iterative equations for the matter part is straightforward, as the corresponding part of the stress-energy tensor does not contain the spin-2 field explicitly. That is, the matter part of the right-hand side of \eqref{eq:ford} is independent of $h^{ab}$ at the lowest order, linear at first order, and so on, making the integration of this part of the equation trivial. The resulting action is obtained as a Taylor series which can be summed. The formal result of this sum is the free matter action expressed in terms of a curved metric, $\mathscr{A}_{\text{M},0}(g)$ with $g^{ab}:=\eta^{ab}+\lambda h^{ab}$ \cite{Deser1970,Padmanabhan2008}. Notice that non-minimal couplings to matter are not ruled out by any consistency condition, so minimal coupling to the physical metric in the resulting matter action is not a necessity in this approach. It is the gravitational self-interacting part of the iterative procedure which has to be handled carefully, and this is the part we are going to work with in the rest of the text.

Concerning this self-interacting part, one could expect that the resulting theory exhibits a nonlinear deformation of the original gauge invariance, which is broken at each stage of the iterative procedure. The search for this symmetry has been a commonly used route to argue that general relativity should be the only consistent self-interacting theory of the Fierz-Pauli field, as the only nonlinear deformation of such linear symmetry is diffeomorphism invariance \cite{Ogievetsky1965,Fang1979,Wald1986} (see also the related discussion in \cite{Ortin2004}). However, here we would like to understand the interplay between the preservation of this symmetry and the iterative self-coupling procedure, instead of taking its existence as an assumption from the beginning.

In the rest of the section we are going to solve the iterative equations \eqref{eq:cons3} step by step. First we are going to solve these equations in the simplest case in which there are no non-minimal couplings. We show then that there is a unique solution of these iterative equations, which corresponds to a selection of a certain value of the parameter $\mathfrak{s}$ in \eqref{eq:free10}. Then we devote the next subsection to understand the role of non-minimal couplings. Their inclusion will permit us to obtain the general solution to the self-coupling problem.

\subsection{Explicit integration and summation of the series \label{sec:summ}}

In this section we are going to see how to manage the infinite set of iterative equations \eqref{eq:cons3} fot the spin-2 field. Let us start with the first-order iterative equation \eqref{eq:ford}. To do that, we are going to evaluate the right-hand side of this equation and then integrate the functional form to obtain the corresponding left-hand side.

The first step is to apply Hilbert's prescription to obtain the source of the equations of motion. To do that we have to extend the action \eqref{eq:free9} to a general curved metric.\footnote{It is in this step where the ambiguities in the addition of non-minimal coupling terms can arise. We will deal with this ambiguity in the following section, thus making here the simplest choice.} Adopting a minimal prescription we can write:
\begin{equation}
\mathscr{A}_0[\gamma]=\frac{1}{4}\int\text{d}\mathscr{V}_\gamma\,M^{ai}_{\mathfrak{s} \ \ bcjk}(\gamma)\nabla'_a h^{bc}\nabla'_i h^{jk},\label{eq:cov3}
\end{equation}
where we have dropped the $\text{G}$ subindex. Here $\nabla'$ is the covariant derivative with respect to $\gamma_{ab}$ and $\text{d}\mathscr{V}_\gamma:=\text{d}^4x\sqrt{-\gamma}$ the corresponding volume element. Notice that we have changed the metric in the argument of the tensor $M^{ai}_{\mathfrak{s}\ \ bcjk}(\eta)$ defined in \eqref{eq:free10}.

We obtain the stress-energy tensor by performing variations on this metric, and then taking the limit back to flat space. Under such a variation, the action \eqref{eq:cov3} changes as:
\begin{align}
\delta\mathscr{A}_0[\gamma]=\frac{1}{4}\int\text{d}^4x\,\delta[\sqrt{-\gamma}\,M^{ai}_{\mathfrak{s} \ \ bcjk}(\gamma)]\nabla'_a h^{bc}\nabla'_i h^{jk}+\nonumber\\
+\frac{1}{2}\int\text{d}^4x\sqrt{-\gamma}\,M^{ai}_{\mathfrak{s} \ \ bcjk}(\gamma)\delta[\nabla'_a h^{bc}]\nabla'_i h^{jk}.\label{eq:cov5}
\end{align}
The first term gives two contributions, one coming from the variation of the determinant and the other from the variation of $M^{ai}_{\mathfrak{s}\ \ bcjk}(\gamma)$. There are two possible ways of dealing with the former. The first one is to notice that the first-order equation must be traceless so the corresponding term is not going to contribute. This observation can be extended to all orders with the following recipe: do not change the measure in the partial actions $\mathscr{A}_n$ when writing them in terms of the auxiliary metric $\gamma_{ab}$. Although a departure from Hilbert's prescription, this alternative procedure leads to a sensible source to be used in the self-coupling procedure when the constraints on the field $h^{ab}$ are taken into account. We shall follow this approach in this section. A second option is to proceed with no previous knowledge of the restrictions on $h^{ab}$ and integrate the contribution coming from the variation of the determinant. The iterative equations \eqref{eq:cons3} are linear, so we only need to add the corresponding contribution obtained this way to the result of the calculations of this section. We will show in the next section what is the result of this procedure. Of course, this is only an operational choice which does not affect the physical results at the end of the day, when the constraints on the field $h^{ab}$ are considered.

Let us now deal with the second term of \eqref{eq:cov5}. There we have the difference of two Levi-civita connections associated with $\gamma_{ab}$ and $\gamma_{ab}+\delta\gamma_{ab}$, respectively. This difference is characterized by the tensor
\begin{equation}
{C'}^b_{ad}:=\frac{1}{2}(\gamma^{be}+\delta\gamma^{be})\nabla'_\mu(\gamma_{\nu\rho}+\delta\gamma_{\nu\rho})D^{\mu\nu\rho}_{\ \ \ \ aed},\label{eq:relcon}
\end{equation}
where
\begin{equation}
D^{\mu\nu\rho}_{\ \ \ \ aed}:=\delta^{\mu }_a\delta^{(\nu}_d\delta^{\rho)}_e+\delta^{\mu }_d\delta^{(\nu}_a\delta^{\rho)}_e-\delta^{\mu }_e\delta^{(\rho}_a\delta^{\nu)}_d,
\end{equation}
(see e.g. \cite{Wald1984}). Then one can see that the variation $\delta[\nabla'_a h^{bc}]$ is given, at first order, by:
\begin{equation}
\delta {C'}^{(b}_{ad}h^{c)d}=\gamma^{e(b}h^{c)d}\nabla'_\mu\delta\gamma_{\nu\rho}D^{\mu\nu\rho}_{\ \ \ \ aed}.\label{eq:cov8}
\end{equation}
The notation $\delta{C'}^a_{bc}$ here means that we only take the terms in \eqref{eq:relcon} which are linear in the variations $\delta\gamma_{ab}$. If we integrate by parts, the contribution of these terms equals to
\begin{align}
-\frac{1}{2}\int\text{d}^4x\,\delta\gamma_{\nu\rho}\gamma^{e(b}D^{\mu\nu\rho}_{\ \ \ \ aed}M^{ai}_{\mathfrak{s} \ \ bcjk}\nabla_\mu[h^{c)d}\nabla_i h^{jk}]=\nonumber\\
=\frac{1}{2}\int\text{d}^4x\,\delta\gamma^{pq}\gamma_{p\nu}\gamma_{q\rho}\gamma^{de}D^{a\nu\rho}_{\ \ \ \ \mu e(b}M^{\mu i}_{\mathfrak{s} \ \ c)djk}\nabla_a(h^{bc}\nabla_i h^{jk}).\label{eq:cov10}
\end{align}
The corresponding source then takes the form:
\begin{align}
T_{pq}:=-\frac{1}{4}\left.\frac{\delta M^{ai}_{\mathfrak{s} \ \ bcjk}}{\delta\gamma^{pq}}\right|_{\gamma\rightarrow\eta}\nabla_a h^{bc}\nabla_i h^{jk}-\nonumber\\
-\frac{1}{2}\eta_{p\nu}\eta_{q\rho}\eta^{de}D^{a\nu\rho}_{\ \ \ \ \mu e(b}M^{\mu i}_{\mathfrak{s} \ \ c)djk}\nabla_a(h^{bc}\nabla_i h^{jk}).\label{eq:cov9}
\end{align}
Notice that this expression contains second derivatives of the spin-2 field. It is important to notice also that it naturally splits into two kinds of terms, proportional to $(\nabla h)^2$ and $\nabla( h\nabla h)$, respectively. As it stands, it is symmetric under the exchanges $p\leftrightarrow q$ and $b\leftrightarrow c$.

The objective now is to find a term in the action $\lambda\mathscr{A}_1$ whose variation with respect to $h^{ab}$ gives the desired source term. The most general expression which contains no more than two derivatives of the spin-2 field can be always written as:
\begin{equation}
\frac{1}{4}\int\text{d}\mathscr{V}_\eta\,N^{ai}_{\ \ \ bcjkpq}(\eta)h^{pq}\nabla_a h^{bc}\nabla_i h^{jk}.\label{eq:cov14}
\end{equation}
Then taking the functional derivative with respect to $h^{ab}$ we obtain
\begin{align}
\frac{1}{4}\int\text{d}\mathscr{V}_\eta\left[N^{ai}_{\ \ \ bcjkpq}(\eta)\nabla_a h^{bc}\nabla_i h^{jk}-\right.\nonumber\\
-\left.2N^{ai}_{\ \ \ pqjkbc}(\eta)\nabla_a(h^{bc}\nabla_ih^{jk})\right]\delta h^{pq}.\label{eq:cov14b}
\end{align}
We get two equations coming from the comparison of the coefficients accompanying the two independent combinations $(\nabla h)^2$ and $\nabla(h\nabla h)$ in both eqs. \eqref{eq:cov9} and \eqref{eq:cov14b}:
\begin{equation}
N^{ai}_{\ \ \ bcjkpq}(\eta)=\left.\frac{\delta M^{ai}_{\mathfrak{s} \ \ bcjk}(\gamma)}{\delta\gamma^{pq}}\right|_{\gamma\rightarrow\eta},\label{eq:nm1a}
\end{equation}
and
\begin{equation}
-N^{ai}_{\ \ \ bcjkpq}(\eta)=\eta_{p\nu}\eta_{q\rho}\eta^{de}D^{a\nu\rho}_{\ \ \ \ \mu e(b}M^{\mu i}_{\mathfrak{s} \ \ c)djk}(\eta).\label{eq:nm2a}
\end{equation}
The first equation provides the form of the first-order action \eqref{eq:cov14}. The second equation then becomes a consistency condition that must be satisfied for the whole procedure to be well-defined:
\begin{equation}
\left.-\frac{\delta M^{ai}_{\mathfrak{s} \ \ pqjk}(\gamma)}{\delta\gamma^{bc}}\right|_{\gamma\rightarrow\eta}=\eta_{p\nu}\eta_{q\rho}\eta^{de}D^{a\nu\rho}_{\ \ \ \ \mu e(b}M^{\mu i}_{\mathfrak{s} \ \ c)djk}(\eta).\label{eq:cov16}
\end{equation}
It is this equation which imposes restrictions to the solutions of the iterative equations which, in fact, select $\mathfrak{s}=1$. To obtain this condition on $\mathfrak{s}$, let us notice that the right-hand side of \eqref{eq:cov16} can be written as
\begin{align}
\eta_{pb}M^{ai}_{\mathfrak{s} \ \ qcjk}(\eta)+\delta^a_b\eta_{p\mu}M^{\mu i}_{\mathfrak{s} \ \ qcjk}(\eta)-\nonumber\\
-\eta_{pb}\eta_{q\mu}\eta^{ad}M^{\mu i}_{\mathfrak{s} \ \ dcjk}(\eta),\label{eq:cov18}
\end{align}
where we must impose a symmetrization under the exchange of indices $p\leftrightarrow q$ and $b\leftrightarrow c$. It is useful to write this expression explicitly by using \eqref{eq:free10}, 
\begin{align}
\mathfrak{s}\,\eta_{pb}\left[\eta_{q(j}\delta^a_{k)}\delta^i_c+\eta_{c(j}\delta^a_{k)}\delta^i_q\right]-\eta_{pb}\eta^{ai}\eta_{q(j}\eta_{k)c}+\nonumber\\
+\mathfrak{s}\,\delta^a_b\Big[\eta_{p(j}\eta_{k)q}\delta^i_c+\eta_{p(j}\eta_{k)c}\delta^i_q\Big]-\delta^a_b\eta_{q(j}\eta_{k)c}\delta^i_p-\nonumber\\
-\mathfrak{s}\left[\eta^{ai}\eta_{pb}\eta_{q(j}\eta_{k)c}+\eta_{pb}\eta_{q(j}\delta^a_{k)}\delta^i_c\right]+\delta^i_q\delta^a_{(j}\eta_{pb}\eta_{k)c},
\end{align}
and symmetrize this equation with respect to $p\leftrightarrow q$, so it can be simplified to:
\begin{align}
\mathfrak{s}\Big[\eta_{pb}\eta_{c(j}\delta^a_{k)}\delta^i_q+\eta_{qb}\eta_{c(j}\delta^a_{k)}\delta^i_p\Big]+\delta^a_b\delta^i_c\eta_{p(j}\eta_{k)q}-\nonumber\\
-\frac{\mathfrak{s}+1}{2}\eta^{ai}\Big[\eta_{pb}\eta_{c(j}\eta_{k)q}+\eta_{qb}\eta_{p(j}\eta_{k)c}\Big]+\nonumber\\
+\frac{\mathfrak{s}-1}{2}\delta^a_b\Big[\eta_{p(j}\eta_{k)c}\delta^i_q+\eta_{q(j}\eta_{k)c}\delta^i_p\Big].
\end{align}
This equation must be compared with the left-hand side of \eqref{eq:cov16} i.e. with
\begin{align}
-\left.\frac{\delta M^{ai}_{\ \ \ pqjk}(\gamma)}{\delta\gamma^{bc}}\right|_{\gamma\rightarrow\eta}=\mathfrak{s}\Big[\eta_{pb}\eta_{c(j}\delta^a_{k)}\delta^i_q+\eta_{qb}\eta_{c(j}\delta^a_{k)}\delta^i_p\Big]+\nonumber\\
+\delta^a_b\delta^i_c\eta_{p(j}\eta_{k)q}-\eta^{ai}\Big[\eta_{pb}\eta_{c(j}\eta_{k)q}+\eta_{qb}\eta_{p(j}\eta_{k)c}\Big],\label{eq:cov17}
\end{align}
which must be still symmetrized under the exchange $b\leftrightarrow c$. A direct comparison of these equations tells us that the only solution of \eqref{eq:cov16} is given by $\mathfrak{s}=1$.

In this way, we have shown how to integrate the first-order iterative equation \eqref{eq:ford}. The result is:
\begin{align}
\mathscr{A}_0+\lambda\mathscr{A}_1+\mathscr{O}(\lambda^2)=\nonumber\\
=\frac{1}{4}\int\text{d}\mathscr{V}_\eta\, M^{ai}_{1 \ \ bcjk}(\eta+\lambda h)\nabla_ah^{bc}\nabla_ih^{jk}+\mathscr{O}(\lambda^2).
\end{align}
Now that we have worked out the first order in detail, the objective is to show that the result which can be anticipated from this order is in fact the correct result. That is, that the total action is:
\begin{equation}
\mathscr{A}=\frac{1}{4}\int\text{d}\mathscr{V}_\eta\, M^{ai}_{1 \ \ bcjk}(\eta+\lambda h)\nabla_ah^{bc}\nabla_ih^{jk}.\label{eq:ansatz}
\end{equation}
So we decompose this ansatz \eqref{eq:ansatz} in partial actions, $\mathscr{A}=\sum_{n=0}^\infty\lambda^n\mathscr{A}_n$, with
\begin{equation}
\mathscr{A}_n=\frac{1}{4n!}\int\text{d}^4x\,\left.\frac{\delta^n M^{ai}_{1 \ \ bcjk}(\gamma)}{\delta\gamma^{pq}\delta\gamma^{st}\dots}\right|_{\gamma\rightarrow\eta}\nabla_ah^{bc}\nabla_ih^{jk}h^{pq}h^{st}\dots,
\end{equation}
and apply the iterative equations \eqref{eq:cons3} to this sequence, finding the consistency condition:
\begin{align}
n\,\eta_{p\nu}\eta_{q\rho}\eta^{de}D^{a\nu\rho}_{\ \ \ \ \mu e(b}\left.\frac{\delta^{n-1} M^{\mu i}_{1 \ \ c)djk}(\gamma)}{\delta\gamma^{st}\dots}\right|_{\gamma\rightarrow\eta}=\nonumber\\
=-\left.\frac{\delta^n M^{ai}_{1 \ \ pqjk}(\gamma)}{\delta\gamma^{bc}\delta\gamma^{st}\dots}\right|_{\gamma\rightarrow\eta}.\label{eq:ccond2a}
\end{align}
Notice the symmetrization of the pair $(b,c)$. To work better with this expression, we can avoid at first to evaluate it in the limit $\gamma\rightarrow\eta$, working thus with the equation
\begin{equation}
n\,\gamma_{p\nu}\gamma_{q\rho}\gamma^{de}D^{a\nu\rho}_{\ \ \ \ \mu e(b}\frac{\delta^{n-1} M^{\mu i}_{1 \ \ c)djk}(\gamma)}{\delta\gamma^{st}\dots}=-\frac{\delta^n M^{ai}_{1 \ \ pqjk}(\gamma)}{\delta\gamma^{bc}\delta\gamma^{st}\dots},\label{eq:ccond2}
\end{equation}
which can be viewed as a differential equation with an initial condition imposed in flat space. In fact, if we drop the indices we can write it schematically as
\begin{equation}
n\,\Theta(\gamma) \frac{\partial^{n-1}M(\gamma)}{\partial \gamma^{n-1}}=-\frac{\partial^{n} M(\gamma)}{\partial \gamma^n},\label{eq:esq2}
\end{equation}
with $\Theta\sim (\gamma)^{-1}$. Up to now, we have shown that \eqref{eq:cov16} is valid, which in this simplified notation becomes
\begin{equation}
\Theta(\gamma) M(\gamma)=-\frac{\partial M(\gamma)}{\partial \gamma}.\label{eq:esq1}
\end{equation}
Thus to show by induction that \eqref{eq:ansatz} is in fact the solution to the iterative problem we only need, as we have already proved that it holds for $n=1$, to show that $\Theta(\gamma)$ verifies the differential equation:
\begin{equation}
\frac{\partial \Theta(\gamma)}{\partial \gamma}=-\Theta^2(\gamma),\label{eq:esq2}
\end{equation}
as it is indeed the case. Coming back to the full equations, one has:
\begin{align}
-\frac{\delta^{n+1}M^{ai}_{1 \ \ pqjk}(\gamma)}{\delta\gamma^{uv}\delta\gamma^{bc}\delta\gamma^{st}\dots}=n\,\gamma_{p\nu}\gamma_{q\rho}\gamma^{de}D^{a\nu\rho}_{\ \ \ \mu e(b}\frac{\delta^{n}M^{\mu i}_{1\ \ c)djk}(\gamma)}{\delta\gamma^{uv}\delta\gamma^{st}\dots}+\nonumber\\
+n\,D^{a\nu\rho}_{\ \ \ \mu e(b}\frac{\delta^{n-1}M^{\mu i}_{1\ \ c)djk}(\gamma)}{\delta\gamma^{st}\dots}\frac{\delta}{\delta\gamma^{uv}}\left(\gamma_{p\nu}\gamma_{q\rho}\gamma^{de}\right)+\nonumber\\
+\{(u,v)\leftrightarrow (b,c)\}.
\end{align}
Then the consistency condition with the induction can be read as:
\begin{align}
D^{a\nu\rho}_{\ \ \ \mu e(b}\frac{\delta^{n-1}M^{\mu i}_{1 \ \ c)djk}(\gamma)}{\delta\gamma^{st}\dots}\frac{\delta}{\delta\gamma^{uv}}\left(\gamma_{p\nu}\gamma_{q\rho}\gamma^{de}\right)+\nonumber\\
+D^{a\nu\rho}_{\ \ \ \mu e(u}\frac{\delta^{n-1}M^{\mu i}_{1 \ \ v)djk}(\gamma)}{\delta\gamma^{st}\dots}\frac{\delta}{\delta\gamma^{bc}}\left(\gamma_{p\nu}\gamma_{q\rho}\gamma^{de}\right)=\nonumber\\
=\frac{1}{n}\gamma_{p\nu}\gamma_{q\rho}\gamma^{de}D^{a\nu\rho}_{\ \ \ \mu e(b}\frac{\delta^{n}M^{\mu i}_{1 \ \ c)djk}(\gamma)}{\delta\gamma^{uv}\delta\gamma^{st}\dots}+\nonumber\\
+\frac{1}{n}\gamma_{p\nu}\gamma_{q\rho}\gamma^{de}D^{a\nu\rho}_{\ \ \ \mu e(u}\frac{\delta^{n}M^{\mu i}_{1 \ \ v)djk}(\gamma)}{\delta\gamma^{bc}\delta\gamma^{st}\dots}.
\end{align}
Because of the symmetrization, we can take only one of the terms in each side of the last equation, thus obtaining the equation:
\begin{align}
D^{a\nu\rho}_{\ \ \ \mu e(b}\frac{\delta^{n-1}M^{\mu i}_{1 \ \ c)djk}(\gamma)}{\delta\gamma^{st}\dots}\frac{\delta}{\delta\gamma^{uv}}\left(\gamma_{p\nu}\gamma_{q\rho}\gamma^{de}\right)=\nonumber\\
=\frac{1}{n}\gamma_{p\nu}\gamma_{q\rho}\gamma^{de}D^{a\nu\rho}_{\ \ \ \mu e(u}\frac{\delta^{n}M^{\mu i}_{1 \ \ v)djk}(\gamma)}{\delta\gamma^{bc}\delta\gamma^{st}\dots}=\nonumber\\
=-\gamma_{p\nu}\gamma_{q\rho}\gamma^{de}D^{a\nu\rho}_{\ \ \ \mu e(u}\gamma_{d\alpha}\gamma_{v)\beta}\gamma^{\gamma\delta}D^{\mu\alpha\beta}_{\ \ \ \theta \delta (b}\frac{\delta^{n-1}M^{\theta i}_{1 \ \ \ c)\gamma jk}(\gamma)}{\delta\gamma^{st}}.
\end{align}
In the last line we have used \eqref{eq:ccond2}. So we arrive at the equation:
\begin{align}
D^{a\nu\rho}_{\ \ \ \ \mu eb}\frac{\delta}{\delta\gamma^{st}}\left(\gamma_{p\nu}\gamma_{q\rho}\gamma^{de}\right)=\nonumber\\
=-\gamma_{p\nu}\gamma_{q\rho}\gamma^{\gamma e}D^{a\nu\rho}_{\ \ \ \ \theta e(s}\gamma_{\gamma\alpha}\gamma_{t)\beta}\gamma^{d\delta}D^{\theta\alpha\beta}_{\ \ \ \ \mu \delta b}\label{eq:appeq},
\end{align}
where we have changed the free indices to avoid potential confusions. This is the equation represented schematically by \eqref{eq:esq2}. The reader can find in appendix \ref{sec:ap1} the demonstration that this algebraic relation is indeed true and, therefore, the induction proof is finished.

Notice that, as the construction of the iterative series relies ultimately in the solution of a system of ordinary differential equations schematically represented by \eqref{eq:esq1}, with an initial condition posed in flat space, the solution is unique. The solution $\mathscr{A}$ is then
\begin{align}
\frac{1}{4}\sum_{n=0}^\infty\frac{\lambda^n}{n!}\int\text{d}\mathscr{V}_\eta\,\left.\frac{\delta^nM^{ai}_{1 \ \ \ bcjk}}{\delta\gamma^{pq}\delta\gamma^{st}\dots}\right|_{\gamma\rightarrow\eta}\nabla_ah^{bc}\nabla_ih^{jk}h^{pq}h^{st}\dots=\nonumber\\
=\frac{1}{4}\int\text{d}\mathscr{V}_\eta\,M^{ai}_{1 \ \ \ bcjk}(\eta+\lambda h)\nabla_ah^{bc}\nabla_ih^{jk}=\nonumber\\
=\frac{1}{4\lambda^2}\int\text{d}\mathscr{V}_\eta\,M^{ai}_{1 \ \ \ bcjk}(g)\nabla_ag^{bc}\nabla_ig^{jk}=\mathscr{A},\label{eq:fres}
\end{align}
where we have defined the field
\begin{equation}
g^{ab}:=\eta^{ab}+\lambda h^{ab}.
\end{equation}
Remember that $\nabla$ is the covariant derivative compatible with $\eta_{ab}$.

\subsection{Non-minimal couplings and surface terms \label{sec:nonmin}}

In this section we shall deal with the effect of allowing contributions to the stress-energy tensor coming from non-minimal couplings or, what is equivalent in this case, covariant surface terms. These terms fully parametrize the ambiguity inherent to the definition of the source in the equations of motion. They must be considered for the sake of completeness when the free action in flat space is generalized to a general metric space in terms of the auxiliary metric $\gamma_{ab}$, remember e.g. \eqref{eq:cov3}.

Non-minimal couplings are defined as scalar quantities which can be written in terms of the auxiliary metric $\gamma_{ab}$ in the procedure above and the spin-2 field, and which vanish in the flat-space limit. The most general form of these terms, as they would be added to \eqref{eq:cov3}, is given by:
\begin{equation}
\int\text{d}\mathscr{V}_\gamma\left[A^i_{\ bcjk}(\gamma,\nabla\gamma)h^{bc}\nabla_ih^{jk}+B_{bcjk}(\gamma,\nabla\gamma)h^{bc}h^{jk}\right].\label{eq:uni2b}
\end{equation}
The first function $A^i_{\ bcjk}(\gamma,\nabla\gamma)$ must be proportional to $\nabla\gamma$, while the second one\linebreak $B_{bcjk}(\gamma,\nabla\gamma)$ to $(\nabla\gamma)^2$ or ${\nabla}^2\gamma$. Using the flat covariant derivative $\nabla$ guarantees that these terms vanish in the flat-space limit. We have also restricted them with the condition of leading to contributions to the stress-energy tensor which are quadratic in the derivatives of the spin-2 field. These contributions are obtained by varying this expression with respect to $\gamma^{ab}$ (after integrating by parts) and then taking the flat-space limit. 

The reader could find strange the form \eqref{eq:uni2b} we associate with non-minimal couplings. While the usual representation uses curvature-related tensor quantities, as the Riemann tensor, constructed from specific combinations of the auxiliary metric and its ordinary derivatives $\partial\gamma$, in \eqref{eq:uni2b} we are using arbitrary scalar combinations of the metric and its covariant derivatives $\nabla\gamma$. To do that we are exploiting the fact that we have a Minkowski reference metric, which permits us to easily construct scalar quantities which contain the ($\eta$-covariant) derivatives of the auxiliary metric. Let us consider as an example the Riemann tensor: given a generic decomposition of a metric $\gamma_{ab}$ in the form $\gamma_{ab}=q_{ab}+\epsilon_{ab}$, one can always write its Riemann tensor, $R^i_{\ bcj}(\gamma)$, as
\begin{equation}
R^i_{\ bcj}(\gamma)=R^i_{\ bcj}(q)+2\bar{\nabla}_{[c} \bar{C}^i_{j]b}
+2\bar{C}^i_{d[c}\bar{C}^d_{j]b}.\label{eq:relriemann}
\end{equation}
In this expression, $\bar{C}^a_{bc}$ is the tensor which characterizes the difference between covariant derivatives with respect to the two metrics $\gamma_{ab}$ and $q_{ab}$, respectively denoted by $\nabla'$ and $\bar{\nabla}$ (see for example~\cite{Wald1984}, Eq. D7 adapted to our sign conventions). Now one can consider the special situation in which $q_{ab}=\eta_{ab}$ to realize that the Riemann tensor of $\gamma_{ab}$ can be written as a particular case of the integrand in the expression \eqref{eq:uni2b}.

With this definition of the possible non-minimal couplings, it is not difficult to realize that the same effect can be reproduced by adding a general covariant surface term instead. This term would have the following form, after writing the original action in terms of the auxiliary metric (i.e. it would be added to \eqref{eq:cov3}):  
\begin{equation}
\int\text{d}\mathscr{V}_\gamma\nabla'_a\left(S^{ia}_{\ bcjk}(\gamma) h^{bc}\nabla'_i h^{jk}\right).\label{eq:uni2a}
\end{equation}
As in the case of non-minimal couplings, this is the most general possible expression containing two covariant derivatives of the spin-2 field. Recall that $\nabla'$ is the covariant derivative associated with $\gamma_{ab}$.

Whether surface terms contribute or not to the stress-energy tensor is a matter of choice. If we first substitute the bulk integral by an integral in the boundary, and then perform variations of $\gamma^{ab}$ but maintaining it fixed on this boundary, one would obtain nothing from this variation. If instead one first performs this same variation, one obtains a local contribution to the stress-energy tensor inside the boundary, which does not change the values of global charges. Which stress-energy tensor is the appropriate one can only be distinguished precisely by gravitational experiments. Without further knowledge this is an ambiguity in the definition of the stress-energy tensor (it is equivalent to the ambiguity exploited in the Belinfante-Rosenfeld prescription).

The question now is whether the results we have obtained in the previous section could change because of the introduction of non-minimal couplings. In other words, we want to know whether there exists a different functional
\begin{equation}
\mathscr{A}'_1:=\frac{1}{4}\int\text{d}\mathscr{V}_\eta\,O^{ai}_{\ \ \ bcjkpq}(\eta)h^{pq}\nabla_ah^{bc}\nabla_ih^{jk},\label{eq:uni1}
\end{equation}
solution up to order $\mathscr{O}(\lambda)$ of the iterative procedure when certain additional terms in the stress-energy tensor are taken into account.

The effect of these surface terms or non-minimal couplings would be to add some terms of the form $\nabla(h\nabla h)$ to the stress-energy tensor. Thus, the iterative equations give us two conditions, analogous to \eqref{eq:nm1a} and \eqref{eq:nm2a}: the first one is directly
\begin{equation}
O^{ai}_{\ \ \ bcjkpq}=\left.\frac{\delta M^{ai}_{\mathfrak{s} \ \ bcjk}}{\delta\gamma^{pq}}\right|_{\gamma\rightarrow\eta},\label{eq:nm1}
\end{equation}
as in the minimal coupling case, while the second one will notice the effect of the surface terms, as it is changed to
\begin{align}
-O^{ai}_{\ \ \ bcjkpq}=\nonumber\\
=\eta_{p\nu}\eta_{q\rho}\eta^{de}D^{a\nu\rho}_{\ \ \ \ \mu e(b}M^{\mu i}_{\mathfrak{s} \ \ c)djk}+\Delta^{ai}_{\ \ \ bcjkpq},\label{eq:nm2}
\end{align}
where the term $\Delta^{ai}_{\ \ \ bcjkpq}$ is the contribution coming from the surface term. 

The second equation must be now understood as the condition which permits us to know what surface term we need in order to make the self-coupling procedure consistent for different values of the parameter $\mathfrak{s}$. That is, the addition of surface terms allows us to find solutions to the problem for $\mathfrak{s}\neq 1$. Now the first equation \eqref{eq:nm1} implies that the solution, if it exists, will be expressable as the first term of a Taylor expansion in $\lambda$ of the free action displaced to $\eta^{ab}+\lambda h^{ab}$, for any value of $\mathfrak{s}$. That is,
\begin{align}
\mathscr{A}_0+\lambda\mathscr{A}_1+\mathscr{O}(\lambda^2)=\nonumber\\
=\frac{1}{4}\int\text{d}\mathscr{V}_\eta\, M^{ai}_{\mathfrak{s} \ \ bcjk}(\eta+\lambda h)\nabla_ah^{bc}\nabla_ih^{jk}+\mathscr{O}(\lambda^2).\label{eq:otra}
\end{align} 
Then, the complete iterative procedure will give place to the complete Taylor series in complete analogy with the minimal coupling case \eqref{eq:fres}.

Let us consider now the issue of the variation of the volume element $\text{d}\mathscr{V}_\gamma$ or, in other words, of the factor $\sqrt{-\gamma}$ in the partial actions $\mathscr{A}_n[\gamma]$. The only difference in the integration of the first-order iterative equation is that the variation of the determinant $\delta\sqrt{-\gamma}$ must be taken into account in \eqref{eq:cov5}. This implies that the measure in \eqref{eq:otra} as well as in the final action would be given by $\text{d}\mathscr{V}_g:=\text{d}\mathscr{V}_\eta\,\kappa$ with $\kappa:=\sqrt{-g}/\sqrt{-\eta}$ instead of $\text{d}\mathscr{V}_\eta$. But the effect of this volume element at each order can be absorbed by non-minimal couplings as we did to obtain the solutions \eqref{eq:otra} with values of the parameter $\mathfrak{s}\neq1$. This means that the general solution to the self-coupling problem is given by:
\begin{align}
\mathscr{A}=\frac{1}{4}\int\text{d}\mathscr{V}_\eta\,\kappa'\, M^{ai}_{\mathfrak{s} \ \ bcjk}(\eta+\lambda h)\nabla_a h^{bc}\nabla_i h^{jk}=
\nonumber\\
=\frac{1}{4\lambda^2}\int\text{d}\mathscr{V}_\eta\,\kappa'\,M^{ai}_{\mathfrak{s} \ \ bcjk}(g)\nabla_a g^{bc}\nabla_i g^{jk}~.\label{eq:acteta}
\end{align} 
This form (specially the fact that only the combination $g^{ab}=\eta^{ab}+\lambda h^{ab}$ appears) was not a logical necessity from the beginning, but the analysis shows that it is actually the result. The factor $\kappa'$ is either $\kappa'=1$ or $\kappa'=\kappa$ depending on the prescription we follow to obtain the source at different orders. By expanding this action with respect to $g^{ab}=\eta^{ab}+\lambda h^{ab}$ in the formalism of \cite{Butcher2009}, one can alternatively see that it indeed leads to a solution to the self-coupling problem with the appropriate quadratic (zeroth order) form for each value of the parameter $\mathfrak{s}$, and evaluate the necessary surface terms in a different way. What at the linear level is a surface term, in the final theory is no longer reducible to a surface term, giving place to a complete $\frak{s}$-parameter family of solutions to the problem. Before ending this section, let us recall that all the solutions we have found are constructed as bimetric theories. Although the final theories all exhibit the tensor $g^{ab}$ in their coupling to matter, the flat metric $\eta^{ab}$ forms also part of the construction. We will discuss later what happens if one decides to eliminate $\eta^{ab}$ to make contact with the standard general relativistic formulations.

\section{The full action: relation with unimodular gravity \label{sec:full}}

In this section we briefly analyze the resulting nonlinear theories which we have obtained as solutions to the self-coupling problem. In particular, we investigate the internal gauge symmetry of these theories, and argue that this feature can be used to distinguish between the different possibilities.

Summarizing, we have found a family of theories \eqref{eq:acteta} which depend on a real parameter $\mathfrak{s}$ and are expressed in terms of the variable $g^{ab}=\eta^{ab}+\lambda h^{ab}$. Moreover, this field $g^{ab}$ is constrained by a finite version of a (possibly) nonlinear equation of the form $f_{ab}\delta g^{ab}=0$. This constraint guarantees that the resulting theory has the same degrees of freedom as the original linear construction of the spin-2 field. Being this a scalar constraint, two options arise: $\eta_{ab}h^{ab}=0$ or $\sqrt{-g}=\sqrt{-\eta}$. The first one is the original constraint imposed in the free-field functional space. However, when considering the self-interacting theory it is natural to expect that a modified nonlinear condition unfolds instead of maintaining the original traceless condition. This is the second case which reduces to the first one at the lowest nontrivial order in the coupling constant $\lambda$. These different selections of the parametric and functional freedom lead to different theories. Each theory has its own peculiarities which one can like or dislike. Let us now discuss these peculiarities case by case. Notice that all of these theories are by construction invariant under general changes of coordinates. However, the amount of internal gauge symmetry that they present can be different.

\subsection{Nonlinear trace theories}

In this section we will consider that the deformation $\sqrt{-g}=\sqrt{-\eta}$ holds. Under this condition, $\kappa'=1$ always. The first useful thing to do is try to express the action of the theory, \eqref{eq:acteta}, in an alternative form. To do that, let us introduce the derivative operator $\tilde{\nabla}$ associated with the field $g^{ab}$ interpreted as a spacetime metric, such that:
\begin{equation}
\tilde{\nabla}_ag^{bc}=0.\label{eq:comp}
\end{equation}
Now we can define a tensor field $C^c_{ab}$ relating the two derivative operators $\tilde{\nabla}$ and $\nabla$.\footnote{The most consistent notation with previous definitions would be $\tilde{C}^c_{ab}$ instead of $C^c_{ab}$. However, here we have chosen the latter notation which simplifies the appearance of the subsequent equations.} We are going to show that the entire action can be written in terms of this tensor field.

If we expand the compatibility condition \eqref{eq:comp} and multiply it by the field $g^{ab}$, one has:
\begin{equation}
g^{ak}\nabla_kg^{bc}=-C^b_{kl}g^{ak}g^{cl}-C^c_{kl}g^{ak}g^{bl}.
\end{equation}
By performing permutations of the free indices of this equation, we can write
\begin{equation}
-2C^c_{kl}g^{ak}g^{bl}=g^{ak}\nabla_kg^{bc}+g^{bk}\nabla_kg^{ac}-g^{ck}\nabla_kg^{ab}.
\end{equation}
Now we can multiply this equation by the inverse matrix $g_{ab}$ (which has nothing to do with the contravariant version of $g^{ab}$ obtained by acting twice with the flat metric $\eta_{ab}$) to solve for $C^c_{ab}$:
\begin{align}
C^c_{ab}=-\frac{1}{2}g_{al}g_{bm}\left(g^{lk}\nabla_kg^{mc}+g^{mk}\nabla_kg^{lc}-g^{ck}\nabla_kg^{lm}\right)=\nonumber\\
=-\frac{1}{2}\left(g_{bm}\nabla_ag^{mc}+g_{al}\nabla_bg^{lc}-g_{al}g_{bm}g^{ck}\nabla_kg^{lm}\right).
\end{align}
Then one has:
\begin{align}
g^{ab}C^i_{ja}C^j_{ib}=\nonumber\\
=\frac{g^{ab}}{4}\left(g_{jl}\nabla_ag^{li}+g_{al}\nabla_jg^{li}-g_{al}g_{jm}g^{ik}\nabla_kg^{lm}\right)\times\nonumber\\
\times\left(g_{bn}\nabla_ig^{nj}+g_{in}\nabla_bg^{nj}-g_{ir}g_{bs}g^{jn}\nabla_ng^{rs}\right)=\nonumber\\
=\frac{1}{4}\left(2g_{ln}\nabla_jg^{li}\nabla_ig^{nj}-g_{ln}g_{jm}g^{ik}\nabla_kg^{lm}\nabla_ig^{nj}\right)=\nonumber\\
=\frac{1}{4}M^{ai}_{1\ \ bcjk}(g)\nabla_ag^{bc}\nabla_ig^{jk}.
\end{align}
This means that we can write the action, at least for  the special case $\mathfrak{s}=1$, as:
\begin{equation}
\frac{1}{4\lambda^2}\int\text{d}\mathscr{V}_\eta\,M^{ai}_{1\ \ bcjk}(g) \nabla_a g^{bc} \nabla_i g^{jk}=\frac{1}{\lambda^2}\int\text{d}\mathscr{V}_\eta\,g^{ab}C^i_{ja}C^j_{ib}.\label{eq:finactuni}
\end{equation}
What is interesting about this expression is that it permits us to connect with the usual geometrical language of general relativity, with $g_{ab}$ playing the role of the spacetime metric. To see that, let us consider the Einstein-Hilbert action which contains the curvature scalar $R$ of a metric $g_{ab}$. As we have already discussed in \ref{sec:nonmin}, if the metric is split as $g^{ab} = \eta^{ab}+\lambda h^{ab}$, the curvature scalar can be written in terms of $\eta$-compatible derivatives $\nabla$ of the field $C^c_{ab}$. Then, we can eliminate a surface term by just realizing \cite{Padmanabhan2010} that
\begin{align}
\frac{2}{\lambda^2}\int \text{d}\mathscr{V}_g\,g^{bj}\left(\nabla_{[c} C^{c}_{j]b} + 
C^{c}_{d[c}C^{d}_{j]b} \right)=\nonumber\\
=\frac{2}{\lambda^2}\int\text{d}\mathscr{V}_\eta \nabla_c(\sqrt{-g}\delta^{[c}_ag^{d]b}C^a_{bd})-
\frac{2}{\lambda^2}\int \text{d}\mathscr{V}_g\,g^{bc}C^{c}_{d[c}C^{d}_{j]b}.
\end{align}
The surface term is 
\begin{equation}
\frac{2}{\lambda^2}\int\text{d}\mathscr{V}_\eta \nabla_c(\sqrt{-g}\delta^{[c}_ag^{d]b}C^a_{bd}),\label{eq:surfterm}
\end{equation}
and the remaining action is precisely
\begin{equation}
-\frac{2}{\lambda^2}\int \text{d}\mathscr{V}_g\,g^{bc}C^{c}_{d[c}C^{d}_{j]b}.\label{eq:alm}
\end{equation}
This action was first written by Rosen in the context of a gravitational theory with a preferred flat background \cite{Rosen1940}. Notice that the expression under the integral sign in this last equation is a scalar under general coordinate transformations, so one does not need to complement it with a surface term to ensure this invariance. On the other hand, it is not a scalar under internal gauge transformations as one would need to add the surface term \eqref{eq:surfterm} to guarantee that. In this work we stick to a bimetric formulation (in the sense of Rosen), as our analysis shows that it is really one of the most important consequences of giving full credit to the self-coupling problem. It is not possible to obtain unimodular gravity (or general relativity, see next section) in a strict sense but Rosen's reformulation of it, when starting from a free theory in flat spacetime and using only flat-spacetime notions. 

The only thing we need to do to make full contact with our action \eqref{eq:finactuni} is to impose the condition on the determinant $\sqrt{-g}=\sqrt{-\eta}$. Under this condition, $\text{d}\mathscr{V}_g=\text{d}\mathscr{V}_\eta$ and $C^b_{bc}=0$ as it can be shown by using a particular Minkowski reference frame:
\begin{equation}
\left.C^b_{bc}\right|_{\mathscr{M}}=-\frac{1}{2}g_{ab}\partial_c g^{ab}=
{1 \over \sqrt{-g}}\partial_c \sqrt{-g}=0.
\end{equation}
So the first term in \eqref{eq:alm} can be dropped and the volume element is replaced by $\text{d}\mathscr{V}_\eta$, making this action completely equivalent to \eqref{eq:finactuni}. Therefore, we have recovered the general relativity action, subject to the determinant restriction. Now it is easy to analyze the internal gauge symmetry of this theory, whose generators will be denoted by $\xi^a$. Its infinitesimal counterpart has exactly the same form as a diffeomorphism,
\begin{align}
\delta_\xi h^{ab}=\mathcal{L}_\xi h^{ab}=\nonumber\\
=-\xi^c \nabla_c g^{ab}+g^{ac} \nabla_c \xi^b+g^{bc} \nabla_c \xi^a~,
\label{eq:defgauge}
\end{align}
with the additional condition of preserving the Minkowski volume element:
\begin{equation}
\nabla_a\xi^a=0.\label{eq:cond1}
\end{equation}
Here $\mathcal{L}_\xi$ is the Lie derivative operator. This condition is nothing but the first one in \eqref{eq:tdiff}. As we have the same number of generators subjected to the same number of restrictions there is no reduction of gauge symmetry, just a deformation.

However, the contrary happens when $\mathfrak{s}\neq1$. One would need to impose additional conditions on the generators to make sure that \eqref{eq:defgauge} is a symmetry of the theory for these values of $\mathfrak{s}\neq1$, which means that the gauge symmetry is reduced (see next section for additional comments).

Independently of this we would have, when $\mathfrak{s}\neq1$, additional restrictions coming from the transverse condition:
\begin{equation}
\nabla_ag^{ab}=\nabla_ah^{ab}=0,\label{eq:donder1}
\end{equation}
which can be alternatively written as
\begin{equation}
C^a_{bc}g^{bc}=0.\label{eq:qharm}
\end{equation}
This means that we will also need to impose a deformation of the second condition in the same equation whose expression can be obtained by imposing $\nabla_b \delta_\xi h^{ab}=0$:
\begin{equation}
\square \xi^b + h^{ac}\nabla_a \nabla_c \xi^b + 2(\nabla_a h^{bc})(\nabla_c \xi^a) + g^{bc} \nabla_c \nabla_a \xi^a=0~.\label{eq:findef2}
\end{equation}
Remember that what singles out the case $\mathfrak{s}=1$ from the other values from the point of view of the internal symmetry is that the transverse condition can be dropped even at the level of the free spin-2 theory, which means that the conditions \eqref{eq:donder1} can be relaxed from the beginning. This corresponds to the situation analyzed by symmetry arguments in \cite{vanderBij1981}, being then our discussion compatible with the content of this work (notice that the transverse condition plays no role in the solution of the iterative equations of the self-coupling problem). This is different from $\mathfrak{s}\neq1$, as in those cases we will always need to impose the transverse condition even in the free theory, which implies that the resulting theory will always present the deformation \eqref{eq:findef2} in order to ensure this condition.

Even if we consider $g^{ab}$ as a metric, we still use the natural volume form in Minkowski spacetime to write the action of the resulting theory, \eqref{eq:finactuni}. As we started from a theory formulated in a fixed background, the Minkowski spacetime, there is no logical reason to expect a perfect decoupling of this background structure in the sense of emergent gravity scenarios \cite{Carlip2013}. In fact, we have seen that the presence of this background is only partially camouflaged in the self-interacting theory when all orders in the interaction are taken into account, but not entirely, as this preferred notion of volume survives. However, in this case this non-perfect decoupling is not an undesirable feature: the existence of a background volume element is a definitory characteristic of unimodular gravity and the ultimate reason which makes this theory avoid the first cosmological constant problem \cite{Bombelli1991,Smolin2009}. The consideration of unimodular gravity as the self-interacting theory of a spin-2 field in Minkowski spacetime thus offers a natural way of understanding this feature which, on the other hand, is rather unnatural from the geometrical point of view.

Concerning the purely geometrical point of view, at this stage one could take one step further. Imagine that instead of writing the action in coordinate independent manner we select a particular Cartesian coordinate system. Then, all the covariant derivatives in the action would be substituted by partial derivatives. If someone gives us this action without informing where does it come from we will not have any way of noticing the existence of an external background. The restricted variation conditions will 
become     
\begin{equation}
\sqrt{-g}=1, \qquad \Gamma^a_{bc}g^{bc}=0~,
\end{equation}
i.e. the unit determinant and the so-called harmonic gauge condition. In this way any reference to a Minkowski background is completely erased. The background is so camouflaged that one can even forget it exists. We could switch to a completely geometrical interpretation of the theory. In this form the theory would have to be interpreted similarly to general relativity but only selecting certain coordinate systems as special. This happens because of the mixing of the external and internal symmetries, both are now one and the same symmetry. One can forget about the harmonic condition and allow for almost generic coordinates in the geometrical description. Then only the determinant of the metric is restricted to be minus one. This is precisely what is standardly considered as unimodular gravity. To make or not this geometrization conceptual jump is however optional, not required by the iterative construction, and most importantly, it is not inconsequential. 

\subsection{Linear trace theories}

Now let us consider what happens if the funcional space is constrained by the condition $\eta_{ab}h^{ab}=0$. As in the previous case it is better to start with the particular value $\mathfrak{s}=1$. The invariance of the traceless condition, $\eta_{ab}\delta_\xi h^{ab}=0$, assumed in this particular construction implies the additional requirement for the generators $\xi^a$:
\begin{equation}
\eta_{ab}g^{ac} \nabla_c \xi^b=0~.
\label{eq:cond2}
\end{equation}
The invariance of the transverse condition, $\nabla_b\delta_\xi h^{ab}=0$ implies in this case 
\begin{equation}
\square \xi^b + h^{ac}\nabla_a \nabla_c \xi^b + 2(\nabla_a h^{bc})(\nabla_c \xi^a) =0~.
\label{eq:cond3}
\end{equation}
Therefore, the gauge symmetry of the original linear spin-2 theory has been deformed but also reduced because the generators $\xi^a$ are subjected to three conditions (the transverse condition $\nabla_a\xi^a=0$ if $\kappa'=1$ or the deformation $\tilde{\nabla}_a\xi^a=0$ if $\kappa'=\kappa$ must also hold) instead of two. A similar thing occurs with the transverse condition and the theories with $\mathfrak{s}\neq 1$ independently. We see then that the constraints \eqref{eq:spin2cons} which define the original functional space of the free theory must be suitably relaxed to guarantee that the conditions which their preservation impose on the generators of the transformation \eqref{eq:defgauge} match with the conditions which guarantee that this transformation is in fact a symmetry of the theory, thus making compatible these kinematical and dynamical aspects of the resulting nonlinear theory. If this is not the case, the gauge symmetry of the theory would be reduced. Whether this reduction of symmetry can be aceptable or not seems to be a matter of taste from the perspective of the self-coupling problem only, although a deeper analysis of these theories probably would reveal potentially observable consequences of this reduction. Notice that in this theory the Minkowski metric appears explicitly in the field equations (something similar occurs in the theory which was found by self-coupling in \cite{Blas2007}).

\section{Relation to previous work \label{sec:prev}}

In this section we are going to schematically describe how the techniques developed in previous sections can be applied to Fierz-Pauli theory. We also discuss some of the seemingly contradictory conclusions in the recent literature. From the perspective of the spin-2 theory, Fierz-Pauli theory constitutes an enlargement of the available functional space in which the field $h^{ab}$ is defined, with a parallel enlargement of the internal gauge symmetry. 
Fierz-Pauli theory~\cite{FierzPauli1939} is defined by the action
\begin{equation}
\mathscr{F}_{0}:=\frac{1}{4}\int\text{d}\mathscr{V}_\eta\,F^{ai}_{\mathfrak{s} \ \ bcjk}(\eta)\nabla_a h^{bc}\nabla_i h^{jk},\label{eq:fp1}
\end{equation}
with
\begin{align}
{F_{\frak{s}}}^{ai}_{\ \ bcjk}(\eta):=M^{ai}_{1 \ \ bcjk}(\eta)-2\delta^a_{(b} \delta^i_{c)} \eta_{jk}+\eta^{ai}\eta_{bc}\eta_{jk}+\nonumber\\
+\frac{1-\frak{s}}{2}\Big[
\eta_{j(b}\delta^a_k\delta^i_{c)}+\eta_{k(b}\delta^a_j\delta^i_{c)}-\delta^a_{(b}  \delta^i_j \eta_{c)k} -\delta^a_{(b}  \delta^i_k \eta_{c)j}\Big].
\nonumber
\label{eq:fp2}
\end{align}
The parameter $\frak{s}$ can acquire any value leaving the theory unchanged because the combination 
\begin{equation}
\left(\delta^a_j\delta^i_b\eta_{ck}-\delta^a_b\delta^i_k\eta_{ck}\right)\nabla_ah^{bc}\nabla_ih^{jk}
\label{eq:commutation}
\end{equation}
is just a surface term and this is why in many places Fierz-Pauli theory is presented as the previous one with $\mathfrak{s}$ set to one. In our discussion we are leaving this term explicit, since we have learnt from the spin-2 case that it could be of importance. This action has the internal gauge symmetry
\begin{equation}
\delta h^{ab}_\xi =\eta^{ac}\nabla_c \xi^b+\eta^{bc}\nabla_c \xi^a,
\label{eq:enlgauge}
\end{equation}
where now the generators are unrestricted. This action is the only quadratic action invariant under these gauge transformations \cite{Padmanabhan2008}. We are not going to need here the explicit form of the tensor $F^{ai}_{\mathfrak{s} \ \ bcjk}(\eta)$ though. The only thing we need to keep in mind is that it can be written in terms of $M^{ai}_{1 \ \ bcjk}(\eta)$ plus additional terms, which are now present because the conditions \eqref{eq:spin2cons} no longer hold. 

We can try to apply now the same self-interacting scheme, but with $F^{ai}_{\mathfrak{s} \ \ bcjk}(\eta)$ instead of $M^{ai}_{\mathfrak{s} \ \ cbjk}(\eta)$, to see whether we are able to obtain general relativity as the outcome. However, this procedure does not work out so straightforwardly in this case. The first evidence of this is that there does not exist any $\mathfrak{s}$ for which the analogue of \eqref{eq:cov16} is true, i.e.
\begin{equation}
\sqrt{-\gamma}\,\gamma_{p\nu}\gamma_{q\rho}\gamma^{de}D^{a\nu\rho}_{\ \ \ \ \mu e(b}{F_{\frak{s}}}^{\mu i}_{\ \ \ c)djk}\neq-\frac{\delta \sqrt{-\gamma}{F_{\frak{s}}}^{ai}_{\ \ pqjk}}{\delta\gamma^{bc}}.\label{eq:fp3}
\end{equation}
A way of realizing this is the following: the right-hand side of this equation contains terms which are proportional to $\gamma_{bc}$, not contracted with $F^{ai}_{\mathfrak{s}\ \ pqjk}$, because of the variation of the determinant. However, the left-hand side of this equation does not contain this kind of terms. Independently of the form of $F^{ai}_{\mathfrak{s} \ \ bcjk}$, for the first term in the left hand side of the previous equation one has e.g.
\begin{align}
\gamma_{p\nu}\gamma_{q\rho}\gamma^{de}D^{a\nu\rho}_{\ \ \ \ \mu eb}=\nonumber\\
=\gamma_{b(p}\delta^a_\mu\delta^d_{q)}+\delta^a_b\gamma_{\mu(p}\delta^d_{q)}-\gamma_{b(p}\gamma_{q)\mu}\gamma^{ad}.
\end{align}
The index $b$ never appears in combination with the free index $c$. The same happens with the second term in (\ref{eq:fp3}). This means that not all the sources are valid for the self-coupling procedure, if we want the final theory to be describable by a Lagrangian theory. In particular, we have seen that the application of the Hilbert prescription to the free Fierz-Pauli action \eqref{eq:fp1} leads to a stress-energy tensor which cannot be derived from an action by performing variations of $h^{ab}$ \cite{Ortin2004}. The requeriment that the equations of motion be derivable from an action cannot be ignored, since it is in the heart of the definition of the iterative self-coupling problem. 

The natural thing to check next is whether it is possible to find a solution of the iterative equations by introducing non-minimal couplings or, what is equivalent, covariant surface terms. In particular, one can ask whether general relativity can appear as a result of such self-coupling scheme. This question is clearly and affirmatively answered in the work by Butcher at al.~\cite{Butcher2009}. To do that, these authors performed a reverse engineering exercise. Let us discuss it briefly here. The Einstein-Hilbert action can be expanded as a series in $\lambda$ using the descomposition $g^{ab}=\eta^{ab}+\lambda h^{ab}$. By construction, this series is a solution of the iterative equations \eqref{eq:cons3}.\footnote{It can be checked that, as long as we are performing a Taylor expansion of a function of the form $F(\eta+\lambda h)$, these conditions are verified.} These authors explicitly show that, to guarantee that the overall procedure makes sense, one needs to accept the following condition. When writing the lowest order $\mathscr{A}_0$ in terms of an auxiliary metric $\gamma_{ab}$ to obtain $\mathscr{A}_0(\gamma)$, this quantity must contain non-minimal couplings as they are necessary to obtain the stress-energy tensor appearing in the lowest-order iterative equation (this happens also for higher orders). The quadratic action $\mathscr{A}_0(\gamma)$, when particularized to Minkowski space, leads precisely to $\mathscr{F}_0$ in \eqref{eq:fp1} with $\frak{s}=1$.

Thus there exists a certain source, obtained through the addition of non-minimal couplings, which permits to recover general relativity as a self-interacting theory of the Fierz-Pauli field. As it happened before with the spin-2 theory, here the non-minimal couplings can be understood as surface terms in the free Lagrangian density. In fact, with the appropriate addition of surface terms all of the actions \eqref{eq:fp1} with an arbitrary value of $\mathfrak{s}$ can be uplifted to nonlinear theories that are solutions of the iterative equations. Another matter is whether these final theories have some internal gauge symmetry or not. What is clear is that only the value $\mathfrak{s}=1$ leads to a theory with an internal symmetry of the form of the usual diffeomorphism invariance.

If we consider the non-tensorial general-relativity action
\begin{equation}
\frac{1}{\lambda^2}\int\text{d}^4x\sqrt{-g}\,g^{ab}(\Gamma^c_{da}\Gamma^d_{cb}-\Gamma^c_{ab}\Gamma^d_{cd}),\label{eq:comp1}
\end{equation}
and perform an expansion in the parameter $\lambda$ with $g^{ab}=\eta^{ab}+\lambda h^{ab}$, we will see that it precisely exhibits a coupling term of the form $h^{ab}S_{ab}$ at first order in $\lambda$. Padmanabhan pointed out the role of this object $S_{ab}$ in any coupling scheme leading to general relativity \cite{Padmanabhan2008}. He showed for instance that this object $S_{ab}$ can be obtained from the quadratic term (zeroth order in $\lambda$) by applying only a half-covariantization scheme which might be regarded at least as unnatural (see appendix A in \cite{Butcher2009} for additional comments on this quantity). Somewhat surprisingly, whereas the quadratic action is tensorial, the first order correction should already be non-tensorial. The variation of this new action with respect $\gamma_{ab}$ might lead in principle to a non-tensorial stress-energy object (though finally this is not the case). Therefore, one could argue, as Padmanabhan, that the construction of general relativity from a self-coupling scheme is somewhat unphysical (only at the end of the iterative procedure one would realize the existence of a surface term allowing the construction of a diffeomorphism-invariant Lagrangian density).

However, in our formulation we always keep track of the flat reference metric. This allows us to construct the tensorial action
\begin{equation}
\frac{1}{\lambda^2}\int\text{d}^4x\sqrt{-g}\,g^{ab}(C^c_{da}C^d_{cb}-C^c_{ab}C^d_{cd}),\label{eq:comp2}
\end{equation}
instead of the non-tensorial action \eqref{eq:comp1}. The cubic term has a form $h^{ab}\bar{S}_{ab}$ where now $\bar{S}_{ab}$ is a proper tensor. Moreover, this object is not and must not be the stress-energy tensor. We have seen that there is a natural definition of $\bar{S}_{ab}$ within the iterative procedure, as the result of the integration of the first-order iterative equation analogue to \eqref{eq:ford}. In fact one of the main differences between the work of Padmanabhan \cite{Padmanabhan2008} and that in here is that we have explicitly performed the integration of the iterative equations. In other words, from the point of view of the self-coupling consistency problem, $\bar{S}_{ab}$ is just a derived quantity and not a fundamental one. One will be led to it by following the equations carefully [recall for example the discussion around \eqref{eq:cov14}]. Concerning this last point, Deser makes a similar comment in his reply to Padmanabhan~\cite{Deser2010}: the only role of $\bar{S}_{ab}$ is to lead to the required source when the variations which respect to $h^{ab}$ are performed, and this is precisely the definition of this quantity.

Concerning the surface term: the Einstein-Hilbert action can be partitioned in a first-derivative action plus a surface term in several ways. If one does not introduce a fiducial background metric, this partition has to be non-tensorial. Instead, by introducing a flat background metric, one discovers a tensorial partition. In our view what is unnatural from the self-coupling program is precisely to forget about the background metric, making an identification of the coordinate invariance and the invariance under gauge transformations. Once one obtains the action \eqref{eq:comp2}, which is a scalar, one would not look for complementing this action with additional surface terms to build the scalar curvature. Only when taking the non-trivial conceptual jump of forgetting about the 
background structure and taking a complete geometrical description in terms of a single metric, one would start worrying about the significance of the surface term and its non-tensorial character. In 
this stage we agree with Padmanabhan's~\cite{Padmanabhan2008} that the surface term of the Einstein-Hilbert action is not naturally obtainable by the self-coupling procedure, one has to add geometrical information.

Butcher et al.~\cite{Butcher2009} say that ``\emph{general relativity cannot be derived from energy-momentum self-coupling the Fierz-Pauli Lagrangian}''. More precisely what they mean is that one cannot use the stress-energy tensor obtained straighforwardly from the Fierz-Pauli Lagrangian density by using a minimal coupling prescription. One has to add specific non-minimal couplings. From reading this paper and Padmanabhan's one ends up with the impression that to obtain general relativity from self-interaction one needs to know somehow the final result, as one needs to make use of curved-spacetime notions. However, here we have shown that non-minimal couplings are encompased by covariant surface terms, which form part of the standard arbitrariness in defining the stress-energy tensor even in flat spacetime. Allowing surface terms one finds a one-parameter familly of solutions to the self-coupling problem. From them, general relativity is selected by requiring the final theory to have the largest possible amount of gauge invariance. Thus the construction only uses concepts based on Poincar\'e-invariant field theory and gauge invariance (although some of the mathematical tools used can be geometrical, as the Hilbert prescription to obtain the stress-energy tensor). It is instructive to notice that the necessity of considering the addition of identically conserved terms to the source one would obtain directly from the free action is not exclusive of gravity, but the same thing happens when considering the case of Yang-Mills theory in the second-order formalism, as it is explicitly written (but at some extent ignored) in \cite{Deser1970}.

All these comments apply to the classic work of Deser \cite{Deser1970,Deser2010}, e.g. the fact that the resulting action will be written in terms of the covariant derivative $\nabla$ with respect to the flat reference metric. The clever choice of independent variables in that work allowed him to lead to completion the iterative procedure in a single step. Precisely, this selection of variables hides the fact that the stress-energy tensor obtained by varying $\eta^{ab}$ is not the one that one would directly obtain from the minimally-coupled Fierz-Pauli theory. That is, Deser's first order formalism naturally selects the specific surface term (or non-minimal coupling) that leads to Einstein equations. The reader should not confuse this surface term at the level of the free theory with the surface term in the Einstein-Hilbert action, which is put by hand. Notice that all the other potential solutions to the self-coupling problem are absent. One could recover them by using additional surface terms in his direct construction. One cannot simply exclude these possibilities from the perspective of logic, but now the arguments in \cite{Deser1970} do not apply as the resulting iterative series would be now infinite. A more general treatment such as the one presented here is needed, in which the use of specific variables is avoided and which permits to handle infinite series to know the nature of these solutions.

In \cite{Deser2010} it is argued that the non-uniqueness inherent to the use of Noether currents in the very definition of the self-coupling problem is harmless. The argument is that these identically conserved terms which appear in the definition of the source can be absorbed in a redefinition of the Fierz-Pauli field $h^{ab}$. If we want to keep us in the linear level, there is only one possible redefinition: shifting $h^{ab}$ by its trace $\eta_{ab}h^{ab}$. This means that one could absorb only certain types of such identically conserved terms. Even if we forget about this, it is difficult to see how this procedure could work as we argue in the following. Let us start with the first-order self-interacting equation
\begin{equation}
O_{abcd}h^{cd}=\lambda T_{ab}(h^{\rho\sigma})+\lambda \Theta_{ab}(h^{\rho\sigma}),\label{eq:1}
\end{equation}
where $O_{abcd}$ is a given differential operator (whose form can be obtained from the action \eqref{eq:fp1}), $T_{ab}(h^{\rho\sigma})$ is the source we want to consider, and $\Theta_{ab}(h^{\rho\sigma})$ an identically conserved tensor constructed from $h^{ab}$. When $\lambda=0$ we recover the free field equations. Now let us construct a different field
\begin{equation}
{h'}^{ab}=h^{ab}+\lambda f^{ab}(h^{\rho\sigma}),
\end{equation}
with $f^{ab}(h^{\rho\sigma})$ an arbitrary function of $h^{ab}$ (which, if we want to keep at the linear level, should be proportional to $\eta^{ab} \eta_{cd}h^{cd}$). In \cite{Deser2010} it is argued that there always exists a choice of $f^{ab}(h^{\rho\sigma})$ such that the field equations (\ref{eq:1}) can be written as
\begin{equation}
O_{abcd}{h'}^{cd}=\lambda T_{ab}(h'^{\rho\sigma}),\label{eq:2}
\end{equation}
thus absorbing the identically conserved term $\Theta_{ab}(h^{\rho\sigma})$. The function $f^{ab}(h^{\rho\sigma})$ is determined by the following equation:
\begin{equation}
O_{abcd}{f}^{cd}(h^{\rho\sigma})=-\Theta_{ab}(h^{\rho\sigma}).\label{eq:3}
\end{equation}
As the operator $O_{abcd}$ satisfies $\nabla^a O_{abcd}=0$ [10], this equation is well posed so it seems that one can conclude that the identically conserved terms can be shifted away. However, one should not forget about the stress-energy tensor, which is not a mere spectator here but explicitly depends in the Fierz-Pauli field $h^{ab}$. Thus, at best, one can get instead of (\ref{eq:2}) an equation of the form
\begin{equation}
O_{abcd}{h'}^{cd}=\lambda T'_{ab}(h'^{\rho\sigma}),\label{eq:4}
\end{equation}
such as
\begin{equation}
T'_{ab}(h'^{\rho\sigma})=T_{ab}(h^{\rho\sigma}).
\end{equation}
This means that the effect of the non-minimal couplings cannot be simply shifted away. One must take them into account, so the non-uniqueness problem remains open until an additional assumption is imposed.

In summary, in the Fierz-Pauli case the self-coupling problem naturally leads to a one-parameter set of solutions that includes general relativity. From this point of view general relativity naturally emerges from the self-coupling of a Poincar\'e-invariant field theory. However, to select general relativity from the other theories one has to require the existence of a maximal gauge symmetry. There seems to be no alternative guiding principle to directly obtain general relativity. Moreover, what is obtained is closer to the bimetric theory of Rosen \cite{Rosen1940}.\footnote{The reader should not confuse the resulting theory with what is usually considered a bimetric theory, as here one of the metrics (the flat reference metric) is not a dynamical entity.} Thus, although both theories are observationally equivalent in standard situations, there can be conceptual differences when considering extreme situations such as spacetime singularities. 

\subsection{On gravitational energy}

One of the aspects which is expected to be addressed by such a reinterpretation in terms of a flat reference metric of the gravitational theories is the issue of the gravitational energy. Also we have some loose ends in our construction (remember the discussion in \ref{sec:matter} concerning the conservation of the traceless source $\tilde{T}_{ab}$) which demand to investigate this issue. Let us first consider in this section the example of general relativity. As we have been discussing, there are two equivalent descriptions of the same system and, within each one, distinct ways of stating conservation principles: on the one hand, the purely geometrical vision in which quantities are covariantly conserved (i.e. with respect to the covariant derivative $\tilde{\nabla}$ associated with the metric $g_{ab}$). On the other hand, we expect the system to possess conserved quantities associated with Poincar\'e invariance. In particular, there will be a stress-energy tensor associated with translation invariance, which will be conserved with respect to the flat derivative operator $\nabla$ (within the space of solutions). In this section we explicitly recall how these two pictures fit (see \cite{Rosen1940} and e.g. \cite{Weinberg1972}).

If we include matter in our considerations, then the resulting action has two parts. It is easy to realize that, by construction, taking the variational derivative with respect to the contravariant version of the auxiliary metric $\gamma_{ab}$ in the matter part  is equivalent to performing the same operation but with respect to $g_{ab}$. However, this it not true within the gravitational part. This means that the total canonical stress-energy tensor is given by
\begin{equation}
\kappa \Theta^{\ a}_{\text{M}\,b}+t^a_{\ b}:=\mathscr{L}_{\text{T}}\delta^a_b-\frac{\delta\mathscr{L}_{\text{T}}}{\delta(\nabla_a\psi^\mu)}\nabla_b\psi^\mu.
\end{equation}
Here $\mathscr{L}_{\text{T}}:=\mathscr{L}_{\text{G}}+\mathscr{L}_{\text{M}}$ is the total Lagrangian density, $\psi^\mu$ again represents all the fields, and $\kappa=\sqrt{-g}/\sqrt{-\eta}$ as defined before. In this equation, $\Theta^{\ a}_{\text{M}\,b}$ gives the same conserved charges as the Hilbert stress-energy tensor for the matter part, while
\begin{equation}
t^a_{\ b}:=\mathscr{L}_{\text{G}}\delta^a_b-\frac{\delta\mathscr{L}_{\text{G}}}{\delta(\nabla_ah^{cd})}\nabla_bh^{cd}
\end{equation}
is a new object which provides the notion of gravitational energy. The factor $\kappa$ in the matter part arises because of the occurrence of $\sqrt{-g}$ instead of $\sqrt{-\eta}$ as the natural measure in the action. We are using the canonical stress-energy tensor instead of the Hilbert prescription just to make better contact with the literature: the only important thing in the following arguments is the divergence of this quantity on solutions, thus using the Hilbert stress-energy tensor would lead exactly to the same calculations and conclusions.

As we have already mentioned, this overall quantity is conserved with respect to $\nabla$ within the space of solutions, that is:
\begin{equation}
\left.\nabla_a(\kappa \Theta^{\ a}_{\text{M}\,b}+t^a_{\ b})\right|_{\mathscr{S}}=0.\label{eq:pcons}
\end{equation}
The subscript $\mathscr{S}$ means that the Euler-Lagrange equations are used. Now it can be shown that this last equation is equivalent to the covariant conservation of the Einstein tensor,
\begin{equation}
\tilde{\nabla}_aG^a_{\ b}=0.\label{eq:covcons3}
\end{equation}
The first step to show this is to use the gravitational equations of motion which, in suitable units, permit us to write \eqref{eq:pcons} as
\begin{equation}
\nabla_a(\kappa G^a_{\ b}+t^a_{\ b})=0.\label{eq:covcons4}
\end{equation}
One only needs to realize that this equation is now a purely geometrical statement (that is, it is solely written in terms of the metric $g_{ab}$) which, in fact, is equivalent to \eqref{eq:covcons3} as it was shown in \cite{Rosen1940}.

This argument can be extended also to unimodular gravity. In this case, the equations of motion would be:
\begin{equation}
R^a_{\ b}-\frac{1}{4}R\delta^a_b=T^{\ a}_{\text{M}\,b}-\frac{1}{4}T_{\text{M}}\delta^a_b,\label{eq:unieqs}
\end{equation}
where now $T^{\ a}_{\text{M}\,b}$ is the Hilbert stress-energy tensor of the matter part. But \eqref{eq:pcons} still holds (with $\kappa=1$), in which we can certainly replace $\Theta^{\ a}_{\text{M}\,b}$ with $T^{\ a}_{\text{M}\,b}$. If we rewrite the left-hand side of \eqref{eq:unieqs} as $G^a_{\ b}+R\delta^a_b/4$, then instead of the identity \eqref{eq:covcons3} we have the condition
\begin{equation}
\nabla_a(R+T_{\text{M}})=0.
\end{equation}
One can see then that the self-coupling of the spin-2 field solves the problem associated with the conservation of the traceless source $\tilde{T}_{ab}$ in \eqref{eq:fans}. Moreover, one can remember this equation as the condition which permits the equations of motion of unimodular gravity to be formally equivalent to those of general relativity, but with a cosmological constant unrelated to the parameters in the action \cite{Ellis2011}. This means that in both cases (general relativity and unimodular gravity), the conservation of the matter stress-energy tensor within the space of solutions, 
\begin{equation}
\tilde{\nabla}_aT^{\ a}_{\text{M}\,b}|_{\mathscr{S}}=0,\label{eq:ordcons}
\end{equation}
is obtained as a consequence of the Poincar\'e invariance of the theory and the resulting nonlinear equations of motion. Put differently, the conservation of the total energy (written in a perfectly tensorial way) in flat spacetime on the one hand \eqref{eq:pcons}, and the conservation of the gravitational and matter parts independently in a curved dynamical space on the other hand, \eqref{eq:covcons3} and \eqref{eq:ordcons} respectively, are two ways of writing the same thing. It is important to notice that in unimodular gravity, \eqref{eq:ordcons} can only be taken as an assumption additional to the field equations \cite{Ellis2011}, but here we have seen that from the perspective of the self-coupling problem it is in fact a necessary consequence of Poincar\'e invariance.

The tensor $t^a_{\ b}$ could be interpreted as the gravitational stress-energy tensor, in the sense that it can be used to evaluate the corresponding conserved charges. However, its interpretation as providing a local notion of energy is still problematic. The stress-energy tensor of the linear theory is not gauge invariant and this fact remains in the final stress-energy tensor: it is not a problem due to self-coupling. In other words, this feature does not appear when one performs the summation of the series but is present at each order, even in the free theory, so it should not be considered as a solid argument against the self-coupling program as Padmanabhan claims \cite{Padmanabhan2008}. In fact, for us the vision is quite the opposite: if one insists in the preservation of the original gauge symmetry, then even the free theory is pushing you towards some kind of geometrical (non-local) interpretation from the beginning and, thus, the lack of this local notion of energy is really natural from the perspective of the self-coupling problem. Notice that even the matter stress-energy tensor $T^{\text{M}}_{ab}$ becomes a gauge non-invariant quantity as a result of the coupling. However, the gauge transformation of $T^{\text{M}}_{ab}$ is of tensorial form while that of $t^a_{\ b}$ is of non-tensorial form: it can always be gauged to zero in a point. For this reason, as opposed to the case in the matter sector, it is not possible to extract any local meaning of energy from $t^a_{\ b}$. The single-metric geometrical interpretation offers a clear explanation of this issue, associating the gauge to zero of the stress-energy tensor to free-fall observers, which do not detect gravity.

The geometrical interpretation offers another way to define a gravitational stress-energy tensor by varying $g^{ab}$ in the action. The gravitational stress-energy tensor is then directly $G_{ab}$ (a similar proposal is argued for in~\cite{Cooperstock2013,Dupre2009}), that is a tensor with respect to changes of coordinates in Minkowski spacetime as well as gauge transformations. In this interpretation and the previous one, one could say that outside matter there is no gravitational energy. However, notice that this does not mean that gravitational waves do not carry energy. Gravitational wave solutions exist in the theory. They are solutions of the equations of motion. Then, the covariant conservation with respect to the metric $g_{ab}$ of the stress-energy tensor \eqref{eq:ordcons} tells us that these waves can imprint energy in the fields acting as detectors. The situation will be reminiscent of action-at-a-distance theories.

The theories of gravitons considered here as the starting points of the self-coupling procedure do not make use of a local meaning of gravitational energy, as the graviton stress-energy tensor is not a gauge-invariant observable. When the original notion of gauge invariance is preserved, this is also reflected in the resulting theories (unimodular gravity and general relativity). A proper local meaning would appear however if gauge invariance is broken. Then $t^a_{\ b}$ will be a perfectly defined stress-energy tensor. In a sense, we could say that the self-interaction procedure shows us two possible alternative routes: a) gauge invariance is preserved and one is pushed towards a complete geometrical interpretation, or b) self-interaction breaks gauge invariance so that the final theory has a proper notion of local energy over the Minkowski background.

\section{Conclusions}


In this work we have discussed Gupta's original program in detail, concerning the possible theories which arise as self-interacting theories of gravitons propagating in Minkowski spacetime. The discussion applies to quantum theories whose low-energy spectrum contains gravitons interacting with matter in a flat background, as long as one accepts that the long wavelength limit is described by a classical, second-order Lagrangian field theory.

We have explicitly solved the infinite set of iterative equations that appears when using a standard formalism based on the tensor field variable $h^{ab}$ for the graviton, thus complementing previous work in the subject concerning finite series which appear when specific variables are considered. To do that we have constructed a proof by induction and found the formal sum of the resulting series, starting from a free field theory with a minimal gauge invariance motivated by the irreducible spin-2 representation of the Poincar\'e group. Finally, we have extended and contrasted our approach with previous discussions in the literature which start instead from Fierz-Pauli theory, which has a larger gauge symmetry. The formalism we have used has permitted us to explicitly show the interplay between internal gauge invariance (a notion which is clearly separated from changes of coordinates) and the self-coupling procedure. 

The main conclusions of our analysis are the following:

\begin{itemize}
\item
One obtains field equations which are equivalent to those of unimodular gravity and general relativity as the only consistent results of the self-coupling of gravitons as long as one requires that the amount of internal gauge symmetry of the linear theory is preserved, although in a deformed version, in the self-interacting theory (considering always theories with up to second derivatives of the fields; beyond that see~\cite{Wald1986}). The distinction between these two cases comes from the amount of gauge symmetry present in the initial linear theory. Beyond the gauge-preservation condition we have explicitly shown that the construction is completely natural from the perspective of flat spacetime and does not need any information related to geometric notions: the non-minimal couplings which are necessary in some cases are nothing but natural surface terms that form part of the standard ambiguities in the definition of the stress-energy tensor in flat spacetime.

If one does not require the preservation of gauge invariance, the self-coupling problem exhibits other solutions. As far as we can see, the self-interacting process itself does not tell us whether gauge invariance should or should not be preserved, thus making this decision an additional input in the construction. In other words, internal gauge invariance is not generally preserved in the self-coupling procedure. This is an interesting point to have in mind when considering emergent theories of gravity. In this kind of constructions it is not difficult to obtain excitations with the same degrees of freedom than those that a graviton would have, but getting the correct nonlinear dynamics for these excitations is still an open problem \cite{Barcelo2005}.

\item
Even if the resulting theories admit a geometrical interpretation (it is within this interpretation that we strictly speak of unimodular gravity and general relativity), they are naturally some kind of bimetric theory, similar to the construction of Rosen \cite{Rosen1940}, and not directly unimodular gravity and general relativity, which from this perpective have forgotten the existence of a flat reference metric. In any case, no observational difference can be extracted while considering weak field situations; when dealing with extreme situations (geometries with horizons, cosmological solutions, etc.) there still might be some way to distinguish the two interpretations because some solutions might be forbidden. Therefore, the structures of unimodular gravity and general relativity do appear naturally without recoursing to curved spacetime notions, but precisely because of this they appear in a form that does not demand a geometrical interpretation in terms of a unique metric. Their form does not demand neither the suplementation of the action with additional surface terms to build the Einstein-Hilbert action. The geometrical interpretation is certainly appealing as, on the one hand, it provides a natural interpretation of the absence of a local meaning for the gravitational energy (a gauge dependent quantity) and, on the other hand, it makes the theory self-contained, with no externally fixed elements. However, here we adhere to Rosen's comment more than 60 years ago \cite{Rosen1940b}: {\em ``Perhaps this} (flat spacetime interpretation) {\em may be regarded by some as a step backward. It should be noted, however, that this geometrization referred to has never been extended satisfactorily to other branches of physics, so that gravitation is trated differently from other phenomena. It is therefore not unreasonable to wonder whether it may not be better to give up the geometrical approach to gravitation for the sake of obtaining a more uniform treatment for all the various fields of force that are to be found in nature.''} 
Then, one would postpone to a later stage the inquire about the very nature of the background vacuum and its interconnections with the rest of the physical system. Let us stress that this does not need to lead to a bimetric theory in the sense of having two dynamical Lorentzian metrics, as even the existence of a preferred flat background can be an effective feature.  

\item
The problem of not having a well-defined local notion of energy for the spin-2 field is already present in the linear theory, so it is not something that emerges due to the self-interactions. On the contrary, the presence of self-interactions solves the problem. The solution presents itself as two mutually exclusive mechanisms. In mechanism a) self-interactions preserve the amount of gauge invariance and also the gauge non-invariance of the graviton stress-energy tensor. This gauge non-invariance finds a satisfactory explanation within the geometrical interpretation. In mechanism b) self-interactions break gauge invariance so that the graviton stress-energy tensor acquires a well-defined meaning. These two alternatives conform with in principle distinct theories.

\item

From the point of view of self-coupling, there is no compulsory reason for the global vacuum energy to gravitate. We have recovered that the minimal gauge construction of the graviton field leads to the structure of unimodular gravity, a theory that differs from general relativity in that only the traceless part of the total stress-energy tensor enters the field equations. Any vacuum-energy contribution in the form of a cosmological constant becomes decoupled from geometry, although the structure of the theory still permits the addition of a cosmological constant as a phenomenological integration constant unrelated to the physical vacuum energy (see e.g.~\cite{Ellis2013} for a discussion). In particular, this means that one does not need to consider curved backgrounds as the starting point of the iterative procedure, instead of Minkowski spacetime, to directly obtain a nonzero cosmological constant as it was done in~\cite{Deser1987}. However, when using the Fierz-Pauli gauge-extended version, the global vacuum energy does appear in the geometrical equations. Thus, although it is interesting that the minimal field-theoretical realization of the concept of graviton leads through the self-coupling procedure to a degravitation of the vacuum energy (an idea which was first considered in \cite{vanderBij1981}), there is no definite answer in this formalism concerning this problem. One would need to consider specific models to find a definite answer; for instance, it would be interesting to study in detail the situation in string theory \cite{Alvarez2005}. For us, the important lesson we can draw from the discussion presented here of the self-coupling problem is that there exists room for a natural solution to the first cosmological constant problem in theories of emergent gravity, along the lines of what is proposed in \cite{Ellis2011,Padmanabhan2014}.  

Notice that, as claimed by Ellis, one can safely admit that observations prioritize unimodular gravity rather than general relativity \cite{Ellis2013}. Whereas in a geometrical interpretation unimodular gravity seems rather unnatural in constrast to general relativity, it is perfectly natural in the field-theoretical approach and, what is probably more interesting, its occurrence is tied up to a non-perfect decoupling from the background structure in the sense of emergent gravity scenarios \cite{Carlip2013}. This may be taken as a hint in favor of the field-theoretical approach to quantum gravity or, in a broader sense, to emergent gravity proposals incorporating in some low-energy regime a non-perfect decoupling from an underlying flat spacetime \cite{Barcelo2011}.

\end{itemize}


\acknowledgments
Financial support was provided by the Spanish MICINN through the projects FIS2011-30145-C03-01 and FIS2011-30145-C03-02 (with FEDER contribution), and by the Junta de Andaluc\'{\i}a through the project FQM219. R.C-R. acknowledges support from CSIC through the JAE-predoc program, co-funded by FSE.


\appendix
\begin{widetext}
\section{An algebraic identity \label{sec:ap1}}

If we evaluate the derivative with respect to the auxiliary metric and forget momentarily about the simmetrization in the pair $(s,t)$, we can write \eqref{eq:appeq} as
\begin{align}
D^{a\nu\rho}_{\ \ \ \ \mu eb}(\gamma_{ps}\gamma_{\nu t}\gamma_{q\rho}\gamma^{de}+\gamma_{p\nu}\gamma_{qs}\gamma_{\rho t}\gamma^{de}-\gamma_{p\nu}\gamma_{q\rho}\delta^d_s\delta^e_t)=\gamma_{p\nu}\gamma_{q\rho}\gamma_{t\beta}\gamma^{d\delta}D^{a\nu\rho}_{\ \ \ \ \theta se}D^{\theta\beta e}_{\ \ \ \ \mu\delta b}.\label{eq:condf}
\end{align}
Of course, this equation would only be valid when the terms obtained under the exchange $s\leftrightarrow t$ are added. In the following we are going to show that this equation holds. The left-hand side is easier to evaluate; it is composed by three terms:
\begin{align}
\frac{1}{2}\gamma_{ps}\gamma_{\nu t}\gamma_{q\rho}\gamma^{de}\left(\delta^a_\mu\delta^\nu_b\delta^\rho_e+\delta^a_\mu\delta^\rho_b\delta^\nu_e+\delta^a_b\delta^\nu_\mu\delta^\rho_e+\delta^a_b\delta^\rho_\mu\delta^\nu_e-\delta^a_e\delta^\nu_b\delta^\rho_\mu-\delta^a_e\delta^\rho_b\delta^\nu_\mu\right)+\nonumber\\
+\frac{1}{2}\gamma_{p\nu}\gamma_{qs}\gamma_{\rho t}\gamma^{de}\left(\delta^a_\mu\delta^\nu_b\delta^\rho_e+\delta^a_\mu\delta^\rho_b\delta^\nu_e+\delta^a_b\delta^\nu_\mu\delta^\rho_e+\delta^a_b\delta^\rho_\mu\delta^\nu_e-\delta^a_e\delta^\nu_b\delta^\rho_\mu-\delta^a_e\delta^\rho_b\delta^\nu_\mu\right)-\nonumber\\
-\frac{1}{2}\gamma_{p\nu}\gamma_{q\rho}\delta^d_s\delta^e_t\left(\delta^a_\mu\delta^\nu_b\delta^\rho_e+\delta^a_\mu\delta^\rho_b\delta^\nu_e+\delta^a_b\delta^\nu_\mu\delta^\rho_e+\delta^a_b\delta^\rho_\mu\delta^\nu_e-\delta^a_e\delta^\nu_b\delta^\rho_\mu-\delta^a_e\delta^\rho_b\delta^\nu_\mu\right).
\end{align}
The first six terms are:
\begin{align}
\gamma_{ps}\gamma_{bt}\delta^a_\mu\delta^d_q+\gamma_{ps}\gamma_{qb}\delta^a_\mu\delta^d_t+\gamma_{ps}\gamma_{t\mu}\delta^a_b\delta^d_q+\gamma_{ps}\gamma_{q\mu}\delta^a_b\delta^d_t-\gamma_{ps}\gamma_{q\mu}\gamma_{bt}\gamma^{ad}-\gamma_{ps}\gamma_{qb}\gamma_{t\mu}\gamma^{ad}.
\end{align}
These are followed by the following six terms:
\begin{align}
\gamma_{pb}\gamma_{qs}\delta^a_\mu\delta^d_t+\gamma_{qs}\gamma_{bt}\delta^a_\mu\delta^d_p+\gamma_{p\mu}\gamma_{qs}\delta^a_b\delta^d_t+\gamma_{qs}\gamma_{t\mu}\delta^a_b\delta^d_p-\gamma_{p\mu}\gamma_{qs}\gamma_{bt}\gamma^{ad}-\gamma_{pb}\gamma_{qs}\gamma_{t\mu}\gamma^{ad}.
\end{align}
The last six terms are:
\begin{align}
-\gamma_{pb}\gamma_{qt}\delta^a_\mu\delta^d_s-\gamma_{pt}\gamma_{qb}\delta^a_\mu\delta^d_s-\gamma_{p\mu}\gamma_{qt}\delta^d_s\delta^a_b-\gamma_{pt}\gamma_{q\mu}\delta^a_b\delta^d_s+\gamma_{p\mu}\gamma_{qb}\delta^d_s\delta^a_t+\gamma_{pb}\gamma_{q\mu}\delta^a_t\delta^d_s.
\end{align}
So we finish with the following expression symmetric in $p\leftrightarrow q$:
\begin{align}
\frac{1}{2}\left(\gamma_{ps}\gamma_{bt}\delta^a_\mu\delta^d_q+\gamma_{qs}\gamma_{bt}\delta^a_\mu\delta^d_p+\right.\gamma_{ps}\gamma_{qb}\delta^a_\mu\delta^d_t+\gamma_{qs}\gamma_{pb}\delta^a_\mu\delta^d_t+\gamma_{ps}\gamma_{t\mu}\delta^a_b\delta^d_q+\gamma_{qs}\gamma_{t\mu}\delta^a_b\delta^d_p+\nonumber\\
+\gamma_{ps}\gamma_{q\mu}\delta^a_b\delta^d_t+\gamma_{qs}\gamma_{p\mu}\delta^a_b\delta^d_t-\gamma_{ps}\gamma_{q\mu}\gamma_{bt}\gamma^{ad}-\gamma_{qs}\gamma_{p\mu}\gamma_{bt}\gamma^{ad}-\gamma_{ps}\gamma_{qb}\gamma_{t\mu}\gamma^{ad}-\gamma_{qs}\gamma_{pb}\gamma_{t\mu}\gamma^{ad}+\nonumber\\
+\gamma_{p\mu}\gamma_{qb}\delta^d_s\delta^a_t+\gamma_{q\mu}\gamma_{pb}\delta^a_t\delta^d_s-\gamma_{pb}\gamma_{qt}\delta^a_\mu\delta^d_s-\gamma_{pt}\gamma_{qb}\delta^a_\mu\delta^d_s\left.-\gamma_{p\mu}\gamma_{qt}\delta^d_s\delta^a_b-\gamma_{q\mu}\gamma_{pt}\delta^a_b\delta^d_s\right).
\end{align}
This expression must be still symmetrized under $s\leftrightarrow t$. When one does this some of the terms cancel,
\begin{equation}
3\leftrightarrow 16,\qquad 4\leftrightarrow 15,\qquad 7\leftrightarrow 18,\qquad 8\leftrightarrow 17;
\end{equation}
leaving the simplified result:
\begin{align}
\frac{1}{2}\left(\gamma_{ps}\gamma_{bt}\delta^a_\mu\delta^d_q+\gamma_{qs}\gamma_{bt}\delta^a_\mu\delta^d_p+\right.\gamma_{ps}\gamma_{t\mu}\delta^a_b\delta^d_q+\gamma_{qs}\gamma_{t\mu}\delta^a_b\delta^d_p-\gamma_{ps}\gamma_{q\mu}\gamma_{bt}\gamma^{ad}-\nonumber\\-\gamma_{qs}\gamma_{p\mu}\gamma_{bt}\gamma^{ad}-\gamma_{ps}\gamma_{qb}\gamma_{t\mu}\gamma^{ad}-\gamma_{qs}\gamma_{pb}\gamma_{t\mu}\gamma^{ad}\left.+\gamma_{p\mu}\gamma_{qb}\delta^d_s\delta^a_t+\gamma_{q\mu}\gamma_{pb}\delta^a_t\delta^d_s\right)_{s\leftrightarrow t}.\label{eq:lres}
\end{align}
This is the equation which we must compare with the right-hand side of \eqref{eq:condf}. In this side, there are 36 terms in total,
\begin{align}
\frac{1}{4}\gamma_{p\nu}\gamma_{q\rho}\gamma_{t\beta}\gamma^{d\delta}\left(\delta^a_\theta\delta^\nu_s\delta^\rho_e+\delta^a_\theta\delta^\rho_s\delta^\nu_e+\delta^a_s\delta^\nu_\theta\delta^\rho_e+\delta^a_s\delta^\rho_\theta\delta^\nu_e-\delta^a_e\delta^\rho_s\delta^\nu_\theta-\delta^a_e\delta^\nu_s\delta^\rho_\theta\right)\times\nonumber\\
\times\left(\delta^\theta_\mu \delta^e_b \delta^\beta_\delta + \delta^\theta_\mu \delta^\beta_b \delta^e_\delta + \delta^\theta_b \delta^e_\mu \delta^\beta_\delta + \delta^\theta_b \delta^\beta_\mu \delta^e_\delta - \delta^\theta_\delta \delta^e_\mu \delta^\beta_b - \delta^\theta_\delta \delta^e_b \delta^\beta_\mu\right).
\end{align}
The 36 terms are given by (the following expression is multiplied by $1/4$):
\begin{align}
\gamma_{ps}\gamma_{qb}\delta^a_\mu\delta^d_t+\gamma_{ps}\gamma_{bt}\delta^a_\mu\delta^d_q+\gamma_{ps}\gamma_{q\mu}\delta^a_b\delta^d_t+\gamma_{ps}\gamma_{t\mu}\delta^a_b\delta^d_q-\gamma_{ps}\gamma_{q\mu}\gamma_{bt}\gamma^{ad}-\gamma_{ps}\gamma_{qb}\gamma_{t\mu}\gamma^{ad}+\nonumber\\
+\gamma_{pb}\gamma_{qs}\delta^a_\mu\delta^d_t+\gamma_{qs}\gamma_{bt}\delta^a_\mu\delta^d_p+\gamma_{p\mu}\gamma_{qs}\delta^a_b\delta^d_t+\gamma_{qs}\gamma_{t\mu}\delta^a_b\delta^d_p-\gamma_{p\mu}\gamma_{qs}\gamma_{bt}\gamma^{ad}-\gamma_{pb}\gamma_{qs}\gamma_{t\mu}\gamma^{ad}+\nonumber\\
+\gamma_{p\mu}\gamma_{qb}\delta^a_s\delta^d_t+\gamma_{p\mu}\gamma_{bt}\delta^a_s\delta^d_q+\gamma_{pb}\gamma_{q\mu}\delta^a_s\delta^d_t+\gamma_{pb}\gamma_{t\mu}\delta^a_s\delta^d_q-\gamma_{q\mu}\gamma_{tb}\delta^a_s\delta^d_p-\gamma_{qb}\gamma_{t\mu}\delta^a_s\delta^d_p+\nonumber\\
+\gamma_{pb}\gamma_{q\mu}\delta^a_s\delta^d_t+\gamma_{q\mu}\gamma_{bt}\delta^a_s\delta^d_p+\gamma_{p\mu}\gamma_{qb}\delta^a_s\delta^d_t+\gamma_{qb}\gamma_{t\mu}\delta^a_s\delta^d_p-\gamma_{p\mu}\gamma_{bt}\delta^a_s\delta^d_q-\gamma_{pb}\gamma_{t\mu}\delta^a_s\delta^d_q-\nonumber\\
-\gamma_{p\mu}\gamma_{qs}\delta^a_b\delta^d_t-\gamma_{p\mu}\gamma_{qs}\gamma_{bt}\gamma^{ad}-\gamma_{pb}\gamma_{qs}\delta^a_\mu\delta^d_t-\gamma_{pb}\gamma_{qs}\gamma_{t\mu}\gamma^{ad}+\gamma_{qs}\gamma_{bt}\delta^a_\mu\gamma^d_p+\gamma_{qs}\gamma_{t\mu}\delta^a_b\delta^d_p-\nonumber\\
-\gamma_{ps}\gamma_{q\mu}\delta^a_b\delta^d_t-\gamma_{ps}\gamma_{q\mu}\gamma_{bt}\gamma^{ad}-\gamma_{ps}\gamma_{qb}\delta^a_\mu\delta^d_t-\gamma_{ps}\gamma_{qb}\gamma_{t\mu}\gamma^{ad}+\gamma_{ps}\gamma_{bt}\delta^a_\mu\delta^d_q+\gamma_{ps}\gamma_{t\mu}\delta^a_b\delta^d_q.
\end{align}
There are several terms which cancels:
\begin{equation}
1-33,\qquad 3-31,\qquad 7-27,\qquad 9-25,\qquad 14-23,\qquad 16-24,\qquad 17-20,\qquad 18-22.
\end{equation}
The 20 restant terms are paired, and they correspond to the 10 terms in \eqref{eq:lres}:
\begin{align}
2+35\sim1,\qquad 4+36\sim3,\qquad5+32\sim5,\qquad6+34\sim7,\qquad8+29\sim2,\nonumber\\
10+30\sim4,\qquad 11+26\sim6,\qquad 12+28\sim8,\qquad13+21\sim9,\qquad 15+19\sim10.
\end{align}
\end{widetext}


\bibliography{spin-2}	

\end{document}